\documentclass[utf8]{frontiersSCNS} % for Science, Engineering and Humanities and Social Sciences articles
\usepackage{url,hyperref,lineno,microtype,subcaption}
\usepackage[onehalfspacing]{setspace}

\newcommand{\aap}{Astronomy \& Astrophysics}
\newcommand{\prl}{Physics Review Letters}
\newcommand{\prd}{Physics Review D}
\newcommand{\solphys}{Solar Physics}
\newcommand{\apjl}{Astrophysical Journal Letters}
\newcommand{\msun}{{\rm M_\odot}}
\newcommand{\rsun}{{\rm R_\odot}}
\newcommand{\lsun}{{\rm L_\odot}}
\newcommand{\zxsun}{{\rm (Z/X)_\odot}}
\newcommand{\tausun}{\tau_\odot}
\newcommand{\xini}{X_{\rm ini}}	
\newcommand{\yini}{Y_{\rm ini}}
\newcommand{\zini}{Z_{\rm ini}}
\newcommand{\ysur}{Y_{\rm s}}
\newcommand{\zsur}{Z_{\rm s}}

\newcommand{\rcz}{R_{\rm CZ}}
\newcommand{\tcz}{T_{\rm CZ}}
\newcommand{\alphamlt}{\alpha_{\rm MLT}}
\newcommand{\phib}{\Phi({\rm ^8B})}
\newcommand{\phibe}{\Phi({\rm ^7Be})}
\newcommand{\phin}{\Phi({\rm ^{13}N})}
\newcommand{\phio}{\Phi({\rm ^{15}O})}
\newcommand{\phif}{\Phi({\rm ^{17}F})}
\newcommand{\phipp}{\Phi({\rm pp})}
\newcommand{\phipep}{\Phi({\rm pep})}
\newcommand{\phihep}{\Phi({\rm hep})}
\newcommand{\phien}{\Phi({\rm eN})}
\newcommand{\phieo}{\Phi({\rm eO})}
\newcommand{\phief}{\Phi({\rm eF})}
\newcommand{\ppS}{{\rm p}({\rm p},e^{+}\nu_e){\rm d}}
\newcommand{\pepS}{{\rm p}({\rm p}e^{-},\nu_e){\rm d}}
\newcommand{\DpS}{{\rm d}({\rm p},\gamma){^3{\rm He}}}
\newcommand{\hethetS}{^{3}{\rm He}({^{3}{\rm He}},2p){^{4}{\rm He}}}
\newcommand{\hetheqS}{^{3}{\rm He}({^{4}{\rm He}},\gamma){^{7}{\rm Be}}}
\newcommand{\beeS}{^{7}{\rm Be}(e^{-},\nu_e){^{7}{\rm Li}}}
\newcommand{\bepS}{^{7}{\rm Be}({\rm p},\gamma){^{8}{\rm B}}}
\newcommand{\hepS}{^3{\rm He}({\rm p},e^{+}\nu_e){^{4}{\rm He}}}
\newcommand{\CpS} {{\rm ^{\rm 12}C(p,\gamma)^{\rm 13}N}}
\newcommand{\NpS} {{\rm ^{\rm 14}N(p,\gamma)^{\rm 15}O}}
\newcommand{\svpp}{\langle\sigma v\rangle_{\rm 11}}
\newcommand{\svtt}{\langle\sigma v\rangle_{\rm 33}}
\newcommand{\svtq}{\langle\sigma v\rangle_{\rm 34}}
\newcommand{\svNp}{\langle\sigma v\rangle_{\rm 114}}
\newcommand{\svCp}{\langle\sigma v\rangle_{\rm 112}}

\newcommand{\svbee}{\langle\sigma v\rangle_{\rm e7}}
\newcommand{\svij}{\langle\sigma v\rangle_{ij}}
\newcommand{\Spp}{S_{11}}
\newcommand{\Stt}{S_{33}}
\newcommand{\Stq}{S_{34}}
\newcommand{\SCp}{S_{112}}
\newcommand{\SNp}{S_{114}}
\newcommand{\Sbep}{S_{17}}
\newcommand{\Sbee}{S_{e7}}
\newcommand{\Sij}{S_{ij}}
\newcommand{\gammapp}{\gamma_{11}}
\newcommand{\gammatt}{\gamma_{33}}
\newcommand{\gammatq}{\gamma_{34}}
\newcommand{\gammaNp}{\gamma_{114}}
\newcommand{\gammaCp}{\gamma_{112}}
\newcommand{\gammabep}{\gamma_{17}}

\newcommand{\gammaij}{\gamma_{ij}}
\newcommand{\tc}{T_{\rm c}}
\newcommand{\xcen}{X_{\rm c}}
\newcommand{\ycen}{Y_{\rm c}}
\newcommand{\zcen}{Z_{\rm c}}

\newcommand{\rhoc}{\rho_{\rm c}}
\newcommand{\Qc}{Q_{\rm c}}
\newcommand{\rneq}{r_{\rm ne}}
\newcommand{\nitrocen}{X_{\rm 14, c}}
\newcommand{\carboneq}{X_{\rm 12}(r_{\rm ne})}
\newcommand{\Rpp}{\lambda_{\rm 11}}

\newcommand{\Rtt}{\lambda_{33}}
\newcommand{\Rtq}{\lambda_{34}}
\newcommand{\Rbee}{\lambda_{e7}}
\newcommand{\Rbep}{\lambda_{17}}
\newcommand{\RNp}{\lambda_{114}}
\newcommand{\RCp}{\lambda_{112}}
\newcommand{\alphaQI}{\alpha(Q,I)}
\newcommand{\Ncn}{\mathcal{N}}
\newcommand{\NcnC}{\mathcal{N}_{\rm c}}
\newcommand{\NcnS}{\mathcal{N}_{\rm s}}
\newcommand{\Dcn}{\Delta}
\newcommand{\DcnC}{\Delta_{\rm c}}
\newcommand{\DcnS}{\Delta_{\rm s}}
\newcommand{\DcnCS}{\Delta^{(\rm cs)}}
\newcommand{\Dcp}{{\mathcal D}_{112}}
\newcommand{\Dcpmed}{{\overline{\mathcal D}}_{112}}
\newcommand{\Dcpmedne}{{\overline {\mathcal D}}_{112}(r_{\rm ne})}

\newcommand{\lsim}{\,\raisebox{-0.13cm}{$\stackrel{\textstyle <}{\textstyle\sim}$}\;}
\newcommand{\gsim}{\,\raisebox{-0.13cm}{$\stackrel{\textstyle>}{\textstyle\sim}$}\,}
\def\keyFont{\fontsize{8}{11}\helveticabold }
\def\firstAuthorLast{F.L.~Villante {et~al.}} %use et al only if is more than 1 author
\def\Authors{Francesco L. Villante\,$^{1,2,*}$ and Aldo Serenelli\,$^{3,4}$}
\def\Address{$^{1}$Dipartimento di Scienze Fisiche e Chimiche, Universit\`a degli Studi dell'Aquila, L'Aquila, Italy\\
$^{2}$Laboratori Nazionali del Gran Sasso (LNGS), Istituto Nazionale di Fisica Nucleare (INFN), Assergi (AQ), Italy\\ 
$^{3}$Institute of Space Sciences (ICE, CSIC), Barcelona, Spain\\
$^{4}$Institut d’Estudis Espacials de Catalunya (IEEC), Barcelona, Spain}
\def\corrAuthor{F.L.~Villante}
\def\corrEmail{villante@lngs.infn.it}

\begin{document}
\onecolumn
\firstpage{1}

\title[The relevance of nuclear reactions for SSMs construction]{The relevance of nuclear reactions for Standard Solar Models construction} 

\author[\firstAuthorLast ]{\Authors} %This field will be automatically populated
\address{} %This field will be automatically populated
\correspondance{} %This field will be automatically populated

\extraAuth{}% If there are more than 1 corresponding author, comment this line and uncomment the next one.
%\extraAuth{corresponding Author2 \\ Laboratory X2, Institute X2, Department X2, Organization X2, Street X2, City X2 , State XX2 (only USA, Canada and Australia), Zip Code2, X2 Country X2, email2@uni2.edu}

\maketitle

\begin{abstract}
The fundamental processes by which nuclear energy is generated in the Sun have been known for many years. However, continuous progress in areas such as neutrino experiments, stellar spectroscopy and helioseismic data and techniques requires ever more accurate and precise determination of nuclear reaction cross sections, a fundamental physical input for solar models. In this work, we review the current status of (standard) solar models and present a detailed discussion on the relevance of nuclear reactions for detailed predictions of solar properties. In addition, we also provide an analytical model that helps understanding the relation between nuclear cross sections, neutrino fluxes and the possibility they offer for determining physical characteristics of the solar interior. The latter is of particular relevance in the context of the conundrum posed by the solar composition, the solar abundance problem, and in the light of the first ever direct detection of solar CN neutrinos recently obtained by the Borexino collaboration. Finally, we present a short list of wishes about the precision with which nuclear reaction rates should be determined to allow for further progress in our understanding of the Sun.

\tiny
\keyFont{ \section{Keywords:} solar physics, solar models, nuclear reactions, nuclear astrophysics, solar neutrino fluxes} %All article types: you may provide up to 8 keywords; at least 5 are mandatory.
\end{abstract}

\section{Introduction}
\label{sec:introduction}
The history of solar models, or standard solar models (SSMs) to be more precise, is formed by three large chapters related to the type of observational and experimental data about the solar interior that existed at any given time. The first part of this history comprises the period over which only neutrino data were available, and it spans about 20 years, from the mid 60s to the early 80s of the past century. During that period, the solar neutrino problem was seen by many as having an origin in the complexities involved in building accurate and precise SSMs, a fundamental part of which is determined by the nuclear reaction rates involved in the generation of the solar nuclear energy. Around the end of that era, the precision of nuclear reaction rates involved in the chains of reactions leading to the production of the different solar neutrino fluxes were on the order to 20 to 30\%. These uncertainties may seem large for present day standards. However, if some faith was put in their accuracy, these uncertainties were small enough that associating the solar neutrino problem to nuclear cross section measurements was highly unlikely \citep{bahcall:1982}. 

In the mid 80s helioseismology, the study of solar oscillations, evolved into a precision branch of solar physics. The sensitivity of the frequency spectrum of these global pressure waves to the details of the interior solar structure allowed their reconstruction by means of inversion methods, see e.g. \cite{deubner:1984, jcd:1985}, in particular of the solar interior sound speed. This (r)evolution peaked during the second half of the 1990s with the establishment of the Global Oscillation Network Group (GONG), a network of six instruments established around the world that carried out resolved radial velocity measurements of the solar surface \citep{harvey:1996} and with the launch of the SoHO satellite, both of which provided rich helioseismic datasets. In turn, this led to determination of the solar interior properties with precision of better than 1\% (and in some cases even an order of magnitude better) \citep{Gough:1996}. These results led to the appearance of a new generation of SSMs \citep{bahcall:1995, jcd:1996}, which were successful in satisfying the tight observational constraints imposed by helioseismology, leaving little room for an astrophysical solution to the solar neutrino problem, as had originally been suggested a few years earlier \citep{elsworth:1990}. Simultaneously, Super-Kamiokande \citep{Fukuda:1998fd,Fukuda:2001nj} led to the precise measurement of $^8$B neutrino flux which, in combination with the results of radiochemical experiments Homestake \citep{Cleveland:1998nv}, Gallex \citep{Hampel:1998xg} and SAGE \citep{Abdurashitov:1999zd} strongly hinted at the existence of solar neutrino oscillations, result confirmed just a few years later by SNO results \citep{Ahmad:2001an,Ahmad:2002jz}. The needs of refined nuclear reaction rates imposed by the type and quality of the new observational and experimental data led to famous revisions of nuclear reaction rates such as NACRE \citep{angulo:1999} and in particular that of Solar Fusion I \citep{Adelberger:1998qm}. In the latter, a critical analysis of the accumulated experimental and theoretical data was performed and consensus values were provided for all relevant nuclear reactions affecting energy generation and neutrino production in the Sun. The improvement in the  uncertainties, in particular, was about a factor of to 2, leading to typical errors around 10\%. Simultaneously, several authors used helioseismic inversion of the solar sound speed to determine, or at least set constraints, on the proton-proton reaction rate, showing that its value had to be within about 15\% of its theoretically determined value \citep{scilla:1998,schlattl:1999,antia:1999,turck:2001,antia:2002}.

The combination of helioseismic constraints and the discovery of neutrino oscillations changed the focus of interest of SSMs. In particular, the accurate and precise determination of neutrino fluxes from individual reactions started playing a fundamental role in the determination of the neutrino oscillation parameters. SSMs became a fundamental source of information, a reference, not just for astrophysics, but for particle physics as well. In 2007, the final and present chapter in this history started when Borexino presented the first measurement of the $^7$Be neutrinos \citep{Borexino:2008a}, originating from a subdominant branch of reactions, the so-called pp-II branch of the pp-chain that accounts for about 10\% of the energy generation of the Sun. Further work by Borexino led to an almost complete characterization of the spectrum of neutrinos from the pp-chain  \citep{Borexino:2018}. Together with the very precise measurement of the $^8$B flux from SNO \citep{SNO:2013} and Super-Kamiokande \citep{SK:2016}, we have come full circle and results from solar neutrino experiments can now be used to learn about the properties of the Sun. This is timely. There is a lingering dispute about which is the detailed chemical composition of the Sun, the solar abundance problem (Sect.~\ref{subsec:compo}), that is intimately linked to the uncertainties in our knowledge of radiative opacities in the solar interior. Solar neutrino data can in principle be used to disentangle this problem \citep{Haxton:2008,Serenelli:2013,Villante:2014}, in particular if the promising results by Borexino on solar CN neutrinos \citep{Borexino:CNO} can be further improved. But progress along this line depends crucially on the accuracy and precision with which nuclear reaction rates are known. The latest compilation, Solar Fusion II \citep{Adelberger:2011}, and subsequent work on specific reactions (Section~\ref{sec:SSM}), show on average a factor of two improvement with respect to the status 10 to 15 years ago, and 5\% uncertainties are nowadays typical. But further work is still needed; uncertainties from nuclear reactions still have a non negligible role in the overall SSMs error budget. 

In Section~\ref{sec:SSM} we summarize the current status of SSMs, review the solar abundance problem, the SSM predictions on the solar neutrino spectrum and the status of nuclear reaction rates affecting model predictions. Section~\ref{Sec:role} presents an analytical formulation of the relation between nuclear reaction rates and solar model properties both for reactions from the pp-chains and CNO-cycles. Section~\ref{sec:numerical} reviews results from numerical SSM calculations, including a detailed assessment of uncertainties and highlighting where progress is most needed, and revises the possibility of using future CN neutrino measurements to determine the solar core C+N abundance.

\section{Standard Solar Models}
\label{sec:SSM}

SSMs are a snapshot in the evolution of a 1\,$\msun$ star, calibrated to match present-day surface properties of the Sun. 
Two basic assumptions in SSM calculations are: 1) after the phase of star formation the Sun was chemically homogenized as a result of the fully convective phase during its contraction along the Hayashi track and before nuclear reactions start altering its initial composition and, 2) at all moments during its evolution up to the present solar age  $\tausun=4.57$\,Gyr mass loss is negligible. The calibration is done by adjusting the mixing length parameter ($\alphamlt$) and the initial helium and metal mass fractions ($\yini$ and $\zini$ respectively) in order to satisfy the constraints imposed by the present-day solar luminosity $\lsun = 3.8418 \times 10^{33} \, {\rm erg}\,{\rm s}^{-1}$, radius $\rsun = 6.9598 \times 10^{10} {\rm cm}$ \citep{Bahcall:2006}, and surface metal to hydrogen abundance ratio $\zxsun$, see sect.\ref{subsec:compo}.  
As a result of this procedure, SSM has no free parameters and completely determines the physical properties of the Sun. It can be then validated (or falsified) by other observational constraints, in particular by those provided by solar neutrino fluxes measurements and helioseismic frequencies determinations. 

The physics input in the SSM is rather simple and it accounts for: convective and radiative transport of energy, chemical evolution driven by nuclear reactions, microscopic diffusion of elements which comprises different processes but among which gravitational settling dominates. Over more than 25 years, since the modern version of the SSM was established with the inclusion of microscopic diffusion \citep{bahcall:1992,jcd:1993}, the continuous improvement of the constitutive physics has brought about the changes and the evolution of SSMs.
In particular, a lot of effort has gone into experimental and theoretical work on nuclear reaction rates. But changes in radiative opacities and the equation of state were also relevant.  
We take here as a reference the results of recent SSM calculations by \cite{Vinyoles:2017}, the so-called Barcelona 2016 (B16, for short) SSMs, which are based on the following state of the art ingredients.
The equation of state is calculated consistently for each of the compositions used in the solar calibrations by using FreeEOS \citep{FreeEOS:2003}.
Atomic radiative opacities are from the Opacity Project (OP) \citep{OP:2005}, complemented at low temperatures with molecular opacities from \cite{Ferguson:2005}. 
Nuclear reaction rates for the pp-chain and CNO-bicycle, which are described in more details in the following section, are from the Solar Fusion II compilation \citep{Adelberger:2011} with important updates for the rates of $\ppS$ \citep{Marcucci:2013, Tognelli:2015,Acharya:2016},  $\bepS$ \citep{Zhang:2015} and $\NpS$ \citep{Marta:2011} reactions. 
Microscopic diffusion coefficients are computed as described in \cite{Thoul:1994}. Convection is treated according to the mixing length theory \citep{Kippenhahn:1990}. The atmosphere is grey and modeled according to a Krishna-Swamy $T-\tau$ relationship \citep{KrishnaSwamy:1966}.

\vspace{0.5cm}

%%%%%%%%%%%%%%%%%%%%%%%%%%%%%%%%%%%%%%%%%%%%%
\subsection{The solar composition problem}\label{subsec:compo}
%%%%%%%%%%%%%%%%%%%%%%%%%%%%%%%%%%%%%%%%%%%%%
The solar surface composition, determined with spectroscopic techniques, is a fundamental input in the construction of SSMs. The development of three dimensional hydrodynamic models of the solar atmosphere, of techniques to study line formation under non-local thermodynamic conditions and the improvement in atomic properties (e.g. transition strengths) have led since 2001 to a complete revision of solar abundances.
Table\,~\ref{tab:compo} lists the abundances determined by different authors for the most relevant metals in solar modeling: GN93~\citep{Grevesse:1993}, GS98~\citep{Grevesse:1998}, AGSS09~\citep{Asplund:2009}, C11~\citep{Caffau:2011} and AGSS15~\citep{Asplund:2015a,Asplund:2015b,Asplund:2015c}. 
%Abundances given in the table come, with the only exception of neon, from spectroscopic analysis of the solar photosphere.

\begin{table}[!ht]
 \begin{center}
\footnotesize
\begin{tabular}{lccccc}
\hline
\hline\noalign{\smallskip}
El. & GN93 & GS98 & AGSS09 & C11 & AGSS15 \\
\noalign{\smallskip}\hline\noalign{\smallskip}
C  & 8.55  & 8.52 & 8.43 & 8.50 & --- \\
N  & 7.97  & 7.92 & 7.83 & 7.86 & --- \\
O  & 8.87  & 8.83 & 8.69 & 8.76 & --- \\
Ne & 8.08  & 8.08 & 7.93 & 8.05 & 7.93 \\
Mg & 7.58  & 7.58 & 7.60 & 7.54 & 7.59 \\
Si & 7.55  & 7.55 & 7.51 & 7.52 & 7.51 \\
S  & 7.33  & 7.33 & 7.13 & 7.16 & 7.13 \\
Fe & 7.50  & 7.50 & 7.50 & 7.52 & 7.47 \\
\noalign{\smallskip}\hline \noalign{\smallskip}
(Z/X)$_\odot$ & 0.0245 & 0.0230 & 0.0180 & 0.0209 & --- \\
\noalign{\smallskip}\hline
\end{tabular}
\end{center}
\caption{\small\em Solar photospheric composition through time and authors for most relevant metals in solar modeling. Abundances are given in the standard astronomical scale  $\epsilon_i=\log_{10}{\left(n_i/n_{\rm H}\right)}+12$, where $n_i$ is the number density of a given atomic species. \label{tab:compo}}  
% Or use
%\vspace*{5cm}  % with the correct table height
\end{table}

Note that only abundances relative to hydrogen can be obtained from spectroscopy because the intensity of spectroscopic lines is measured relative to a continuum that is determined by the hydrogen abundance in the solar atmosphere. The last row in the table gives the total photospheric present-day metal-to-hydrogen ratio $\zxsun$ and it is the quantity used as observational constraint to construct a solar model. In fact, the solar composition set used in solar models determines not only $\zxsun$ but also the relative abundances of metals in the models. In this sense, $\zini$ acts as a normalization factor that, together with $\yini$ and the relation $\xini + \yini + \zini = 1$, determines completely the initial composition of the model.

There is no complete agreement among authors, and some controversy still remains as to what the best values for the new spectroscopic abundances are. %see e.g. \cite{Asplund:2009,Caffau:2011}. 
However, there is consensus in that all determinations of the solar metallicity based on the new generation of spectroscopic studies yield a solar metallicity  lower than older spectroscopic results \citep{Grevesse:1993,Grevesse:1998}, in particular for the volatile and most abundant C, N, and O.
For refractories elements, like Fe, Si, Mg and S that have important role in solar modeling being important contributors to the radiative opacity, meteorites offer a very valuable alternative method  (see e.g. \cite{Lodders:2009}) and, in fact, elemental abundances determined from meteorites have been historically more robust than spectroscopic ones. 

Considering that uncertainties in element abundances are difficult to quantify, it has become customary to consider two canonical sets of abundances to which we refer to as high metallicity (HZ) and low metallicity (LZ) solar admixtures, see e.g. \cite{Serenelli:2011,Vinyoles:2017} as reference assumptions for SSM calculations. These are obtained by using the photospheric (volatiles) + meteoritic (refractories) abundances from GS98 and AGSS09 respectively, and are reported in Table \ref{tab:compo2}. In the last column, we give the fractional differences $\delta z_{i} \equiv z_{i}^{\rm HZ} /z_{i}^{\rm LZ} - 1$ where $z_{i} \equiv Z_{i}/X$ is the ratio of the $i-$element abundance with that of hydrogen, to facilitate comparison among the two admixtures.
Even if GS98 abundances are presumably surpassed by the more recent determinations, they are still considered as a valid option to construct solar models because they lead to a temperature stratification that well reproduces the helioseismic constraints.

\begin{table}[!ht]
\footnotesize
\begin{center}
\begin{tabular}{l c c c}
\hline
\hline\noalign{\smallskip}
 El. &  High-Z (HZ) & Low-Z (LZ) & $\delta z_{i}$\\
\noalign{\smallskip}\hline\noalign{\smallskip}
 C & $8.52  \pm 0.06$ & $8.43 \pm 0.05$ & 0.23 \\
 N & $7.92  \pm 0.06$ & $7.83 \pm 0.05$ & 0.23\\
 O & $8.83  \pm 0.06$ & $8.69 \pm 0.05$ & 0.38\\ 
 Ne & $8.08 \pm 0.06$ & $7.93 \pm 0.10$ & 0.41\\
 Mg & $7.58 \pm 0.01$ & $7.53 \pm 0.01$ & 0.12\\
 Si & $7.56 \pm 0.01$ & $7.51 \pm 0.01$ & 0.12\\
 S  & $7.20 \pm 0.06$ & $7.15 \pm 0.02$ & 0.12\\
 Ar & $6.40 \pm 0.06$ & $6.40 \pm 0.13$ & 0.00 \\
 Fe & $7.50 \pm 0.01$ & $7.45 \pm 0.01$ & 0.12\\
\noalign{\smallskip}\hline\noalign{\smallskip}
$\zxsun$ & 0.02292 & 0.01780 & 0.29 \\ 
\noalign{\smallskip}\hline\noalign{\smallskip}
\end{tabular}
\end{center}
\caption{\small\em The two canonical HZ and LZ solar mixtures  given as $\epsilon_i=\log_{10}{\left(n_i/n_{\rm H}\right)}+12$. 
The two compilations are obtained by using the photospheric (volatiles) + meteoritic (refractories) abundances from GS98 and AGSS09 respectively, and correpond to the admixture labelled as GS98 and AGSS09met in \cite{Vinyoles:2017}.
\label{tab:compo2}}
\end{table}

\begin{figure}[!t]
\begin{center}
\includegraphics[height=6.6cm]{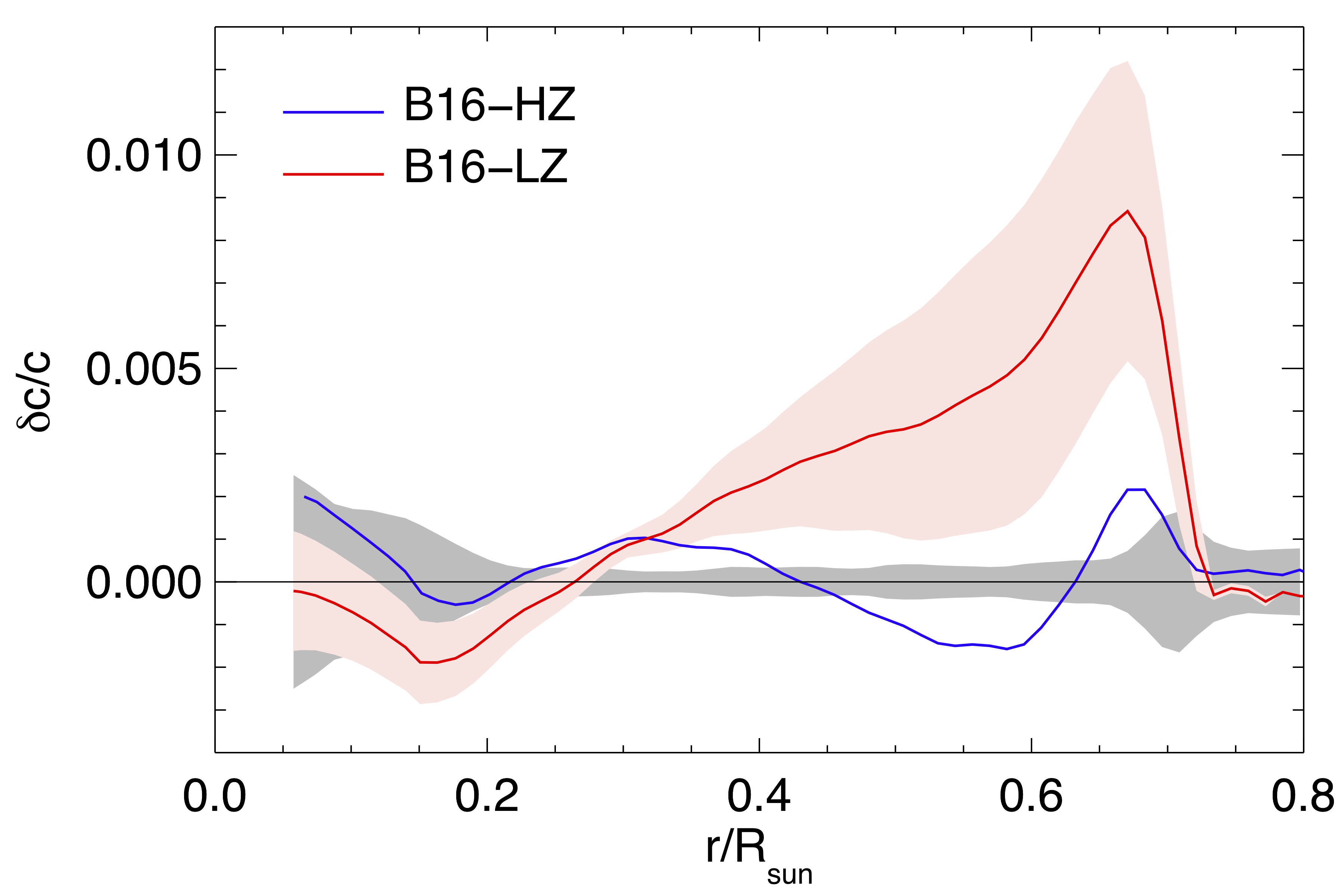}
\caption{\small\em Fractional sound speed difference in the sense $\delta {\rm c /c = (c_\odot - c_{\rm mod})/c_{\rm mod}}$. Grey shaded regions corresponds to errors from helioseismic inversion procedure. Red shaded region corresponds to uncertainties in SSM predictions which we chose to plot around the B16-LZ central value (solid red line). An equivalent relative error band holds around the central value of the B16-HZ central value (solid blue line) which we do not plot for the sake of clarity.
\label{fig:sound}}
\end{center}
\end{figure}

This can be better appreciated by considering Tab.\ref{tab:ssmres} and Fig.\ref{fig:sound} where we compare theoretical predictions of SSMs implementing HZ and LZ surface composition with helioseismic determinations of the surface helium abundance $\ysur$, of the convective envelope depth $\rcz$ and the solar sound speed $c_{\odot}(r)$.
We see that solar models implementing the LZ abundances fail to reproduce all helioseismic probes of solar properties. This disagreement constitutes the so-called \emph{solar abundance problem} \citep{Basu:2004,Bahcall:2005,Delahaye:2006} that has defied a complete solution. 
All proposed modifications to physical processes in SSMs offer, at best, only partial improvements in some helioseismic probes (e.g. \cite{Guzik:2005,castro:2007,Basu:2007,Guzik:2010,Serenelli:2011}). An alternative possibility is to consider modifications to the physical inputs of SSMs at the level of the constitutive physics, radiative opacities in particular.
The effective opacity profile in the solar interior results from the combination of the reigning thermodynamic conditions, including composition, and the atomic opacity calculations at hand. Early works \citep{Bahcall:2005b,Montalban:2004} already suggested that a localized increase in opacities could solve or, at least, alleviate the disagreement of low-Z solar models with helioseismology.  \cite{ChristensenDalsgaard:2009} and \cite{Villante:2010opa} have concluded that a tilted increase in radiative opacities, with a few percent increase in the solar core and a larger (15-20\%) increase at the base of the convective envelope could lead to low-Z SSMs that would satisfy helioseismic probes equally as well as SSMs based on the older, higher, metallicities.

\begin{table}[!ht]
\footnotesize
\begin{center}
%\centering
%\setlength{\tabcolsep}{2.5pt}
\begin{tabular}{lccc}
\hline
\hline\noalign{\smallskip}
Qnt. & B16-HZ & B16-LZ & Solar \\
\noalign{\smallskip}\hline\noalign{\smallskip}
$\ysur$& $0.2426 \pm 0.0059 $&$0.2317 \pm 0.0059 $& $0.2485 \pm 0.0035$ \\
$\rcz/\rsun$& $0.7116 \pm 0.0048$ & $0.7223  \pm 0.0053$ & $0.713 \pm 0.001 $\\
$\langle\delta c/c \rangle$& $0.0005 ^{+0.0006}_{-0.0002}$ &$0.0021 \pm 0.001$ & - \\ \hline
$\alphamlt$ & $2.18 \pm 0.05$ &$2.11 \pm 0.05 $& - \\
$\yini$& $0.2718 \pm 0.0056 $ &$0.2613 \pm 0.0055 $& - \\
$\zini$ & $0.0187 \pm 0.0013$ &$0.0149 \pm 0.0009 $& - \\
$\zsur$& $0.0170 \pm 0.0012 $&$0.0134 \pm 0.0008 $& - \\
$\ycen$& $0.6328 \pm 0.0053 $ &$0.6217 \pm 0.0062 $& -\\
$\zcen$& $0.0200 \pm 0.0014 $ &$0.0159 \pm  0.0010 $& - \\ 
\noalign{\smallskip}\hline\noalign{\smallskip}
\end{tabular}
\end{center}
\caption{\small\em
Main characteristics of SSMs with different surface composition \citep{Vinyoles:2017}. The observational values for $\ysur$ and $\rcz$ are taken from \cite{Basu:2004} and \cite{Basu:1997}, respectively. The quantity $\delta {\rm c /c = (c_\odot - c_{\rm mod})/c_{\rm mod}}$ is the fractional difference between sound speed helioseismic determination and model prediction.}
\label{tab:ssmres}
\end{table}

Recent years have seen a surge of activity in theoretical calculations of atomic radiative opacities. Updated calculations \citep{OP:2005} by the Opacity Project have led the way, followed by OPAS \citep{Blancard:2012,mondet:2015}, STAR \citep{krief:2016b} and a new version of OPLIB, the opacities from Los Alamos \citep{Colgan:2016}. For conditions in solar interiors, all theoretical opacities agree with each other within few \%.
Interestingly, \cite{Bailey:2015} have presented the first ever measurement of opacity under conditions very close to those at the bottom of the solar convective envelope. While the experiment has been carried out only for iron, their conclusion is that all theoretical calculations predict a too low Rosseland mean opacity, at a level of $7\pm 4\%$, for the temperature and density combinations realized in the experiment. Further experimental work on chromium and nickel opacities \citep{nagayama:2019} helps to evaluate discrepancies between experimental and theoretical results on iron opacity. Results point towards shortcomings that affect models, particularly in the case of open electronic L-shell configurations such as is present in iron at the base of the convective envelope. Also, the disagreement between theoretical and measured line shapes for the three elements indicates shortcomings in the theoretical understanding of atomic interaction with the plasma. On the other hand, the results also indicate that the quasicontinuum opacity determined experimentally agrees well with the chromium and nickel experiments, contrary to results from the iron experiment.  However, the chromium and nickel experiments were carried out at lower temperatures than those used in the original iron experiment, which suggests that the problem of missing quasicontinuum opacity might have an unknown temperature dependence, or that a systematic error affected the high temperature iron measurements.
Moreover, \cite{Krief:2016} in a recent theoretical analysis of line broadening modeling in opacity calculations, have found that uncertainties linked to this are larger at the base of the convective envelope than in the core. 
These arguments suggest that opacity calculations are more accurate in the solar core than in the region around the base of the convective envelope.
To take this into account, opacity uncertainty was modelled in B16-SSM calculations in terms of two parameters, $\kappa_a$ and $\kappa_b$, that can change both the scale and the temperature dependence of opacity according to $\delta k(T) = \kappa_a + (\kappa_b/\Delta) \log(T/\tc)$, where $\delta \kappa$ is the fractional opacity variation, $\Delta = \log(\tc/\tcz)$, $\tc = 15.6 \times
10^6\, {\rm K}$ and $\tcz = 2.3 \times 10^6\, {\rm K}$ are the temperatures at the solar center and at the bottom of the
convective zone, respectively. 
The parameters $\kappa_a$ and $\kappa_b$
have been treated as independent random variables with mean
equal to zero and dispersions $\sigma_a = 2\%$ and $\sigma_b = 6.7\%$, corresponding to opacity uncertainty $\sigma_{\rm in} =\sigma_a = 2\% $ at the solar center and $\sigma_{\rm out} =(\sigma_a^2 + \sigma_b^2)^{1/2} = 7\%$ at the base of the convective region.

\vspace{1.5cm}

\subsection{Nuclear reactions in the Sun}
\label{subsec:nuclear}
 
The overall effect of nuclear reactions in the Sun, as in any  other star in hydrogen burning stage, is the conversion:
\begin{equation}
4\,{\rm p} + 2\,e^{-} \to {^{4}{\rm He}} + 2\, \nu_{\rm e} 
\end{equation}
with the production of a fixed amount of energy $Q = 4\, m_{\rm p} + 2 m_{\rm e} - m_{^{4}{\rm He}} = 26.7\,{\rm MeV}$ per synthesized $^{4}{\rm He}$ nucleus. Most of this energy is released in the solar plasma and slowly diffuses toward the solar surface supporting the radiative luminosity of the Sun. A small fraction of it, 
that depends on the specific channel by which hydrogen burning proceeds, is emitted in neutrinos. According to SSM calculations, the two neutrinos carry away about $0.6\,{\rm MeV}$ on the average.

\begin{figure}[!t]
\begin{center}
%\resizebox{0.44\textwidth}{!}{\includegraphics{figure3.eps}}
\includegraphics[width=\columnwidth]{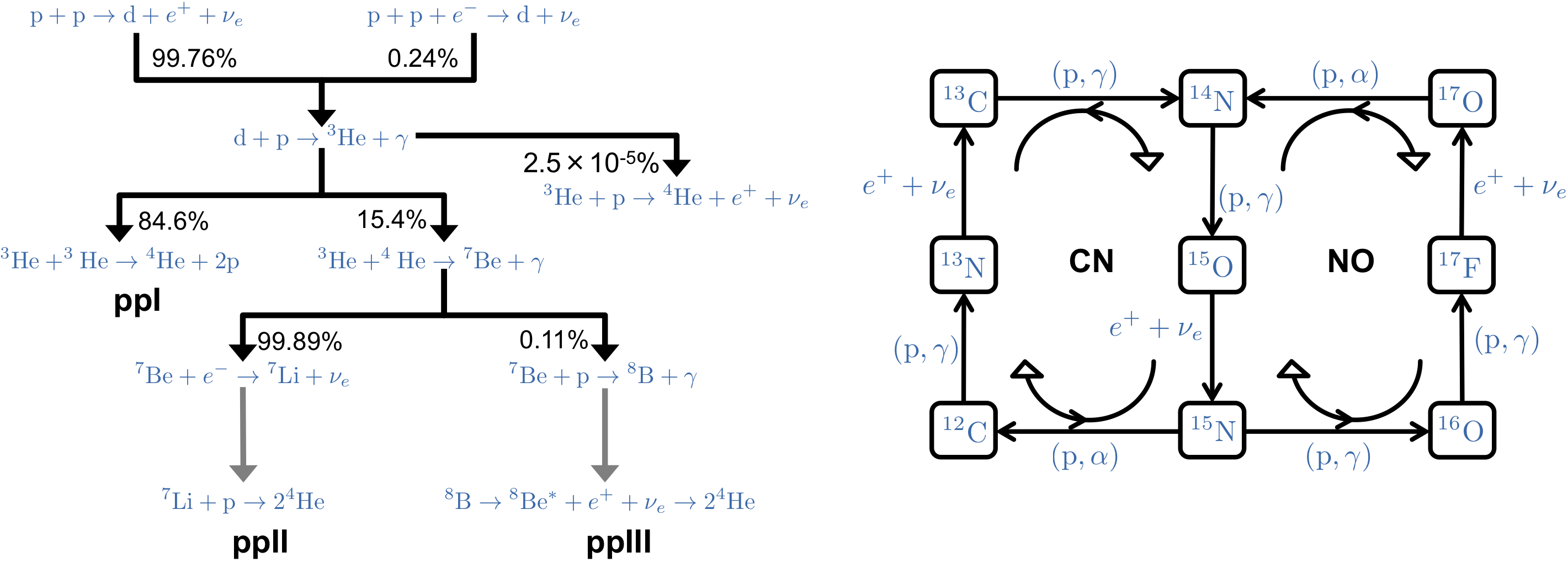}
\caption{\small\em \texttt{\rm Left Panel}: The pp-chain;
\texttt{\rm Right Panel}: The CNO-bicycle.
\label{fig:PPandCNO}}
\end{center}
\end{figure}

\begin{figure}[!t]
\begin{center}
%\resizebox{0.44\textwidth}{!}{\includegraphics{figure3.eps}}
%\includegraphics[width=5.8cm,height=4.5cm]{PhysB16.pdf}
\includegraphics[width=5.8cm,height=4.5cm]{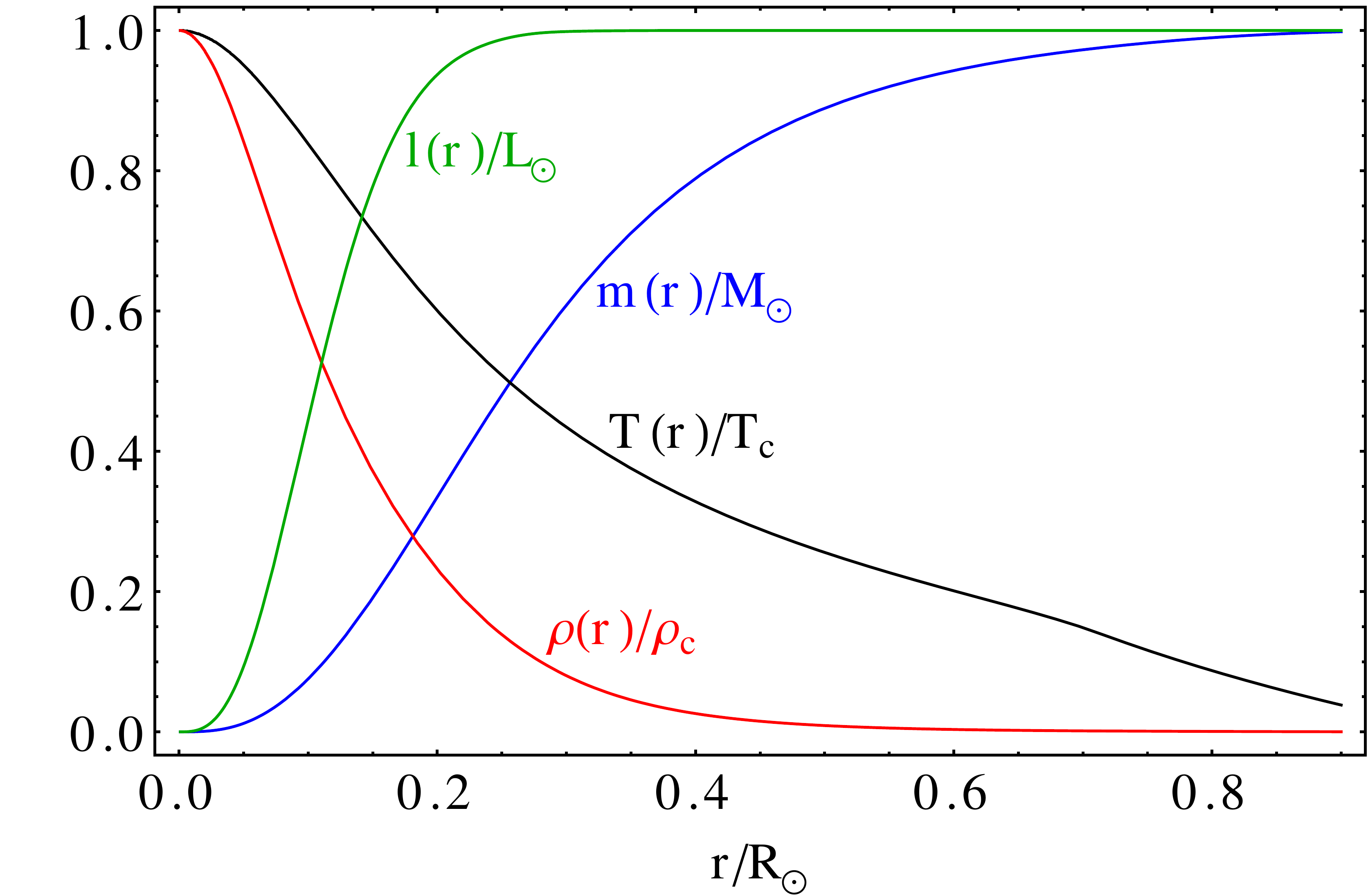}
\includegraphics[width=5.8cm,height=4.5cm]{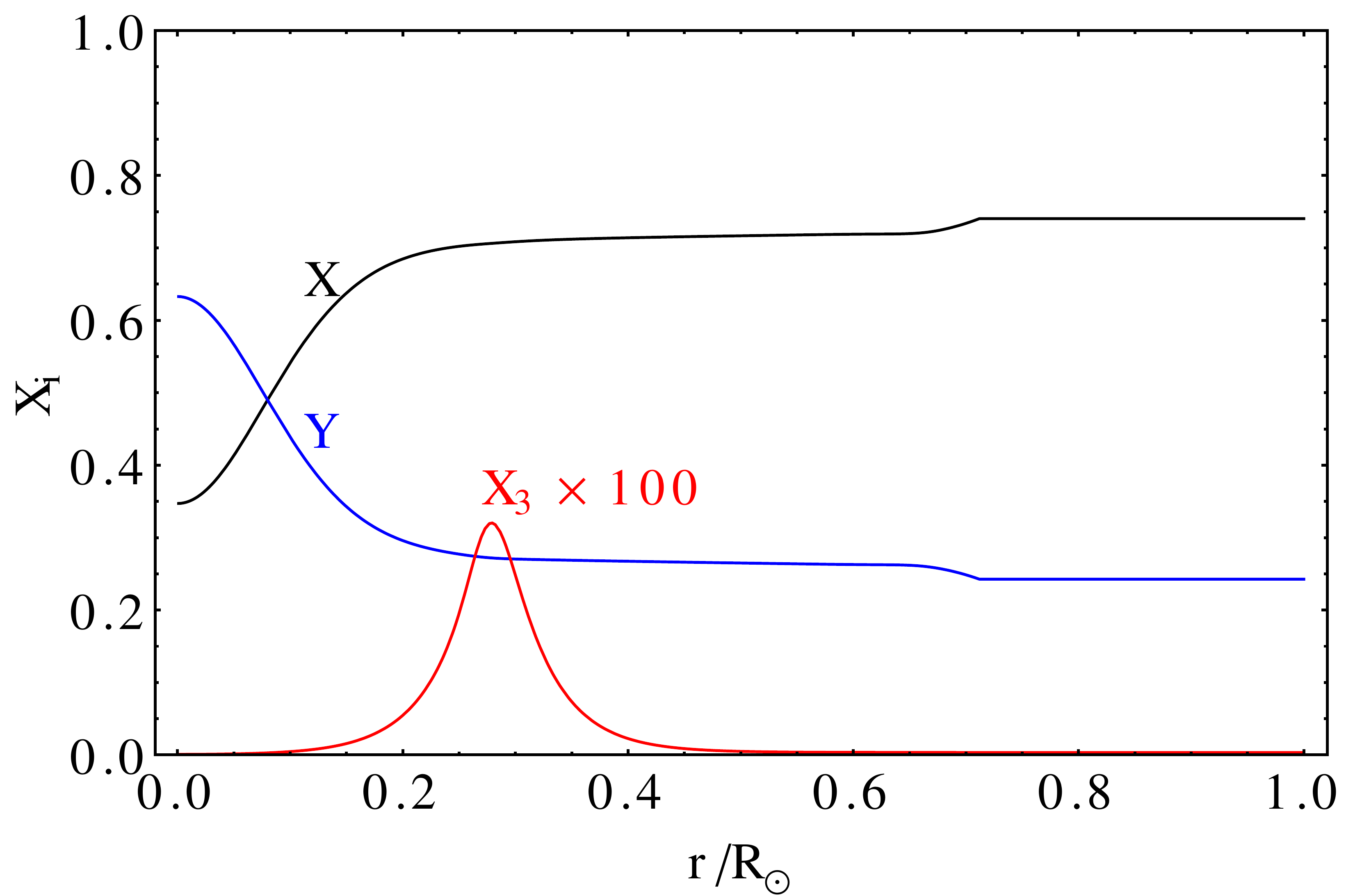}
\includegraphics[width=5.8cm,height=4.5cm]{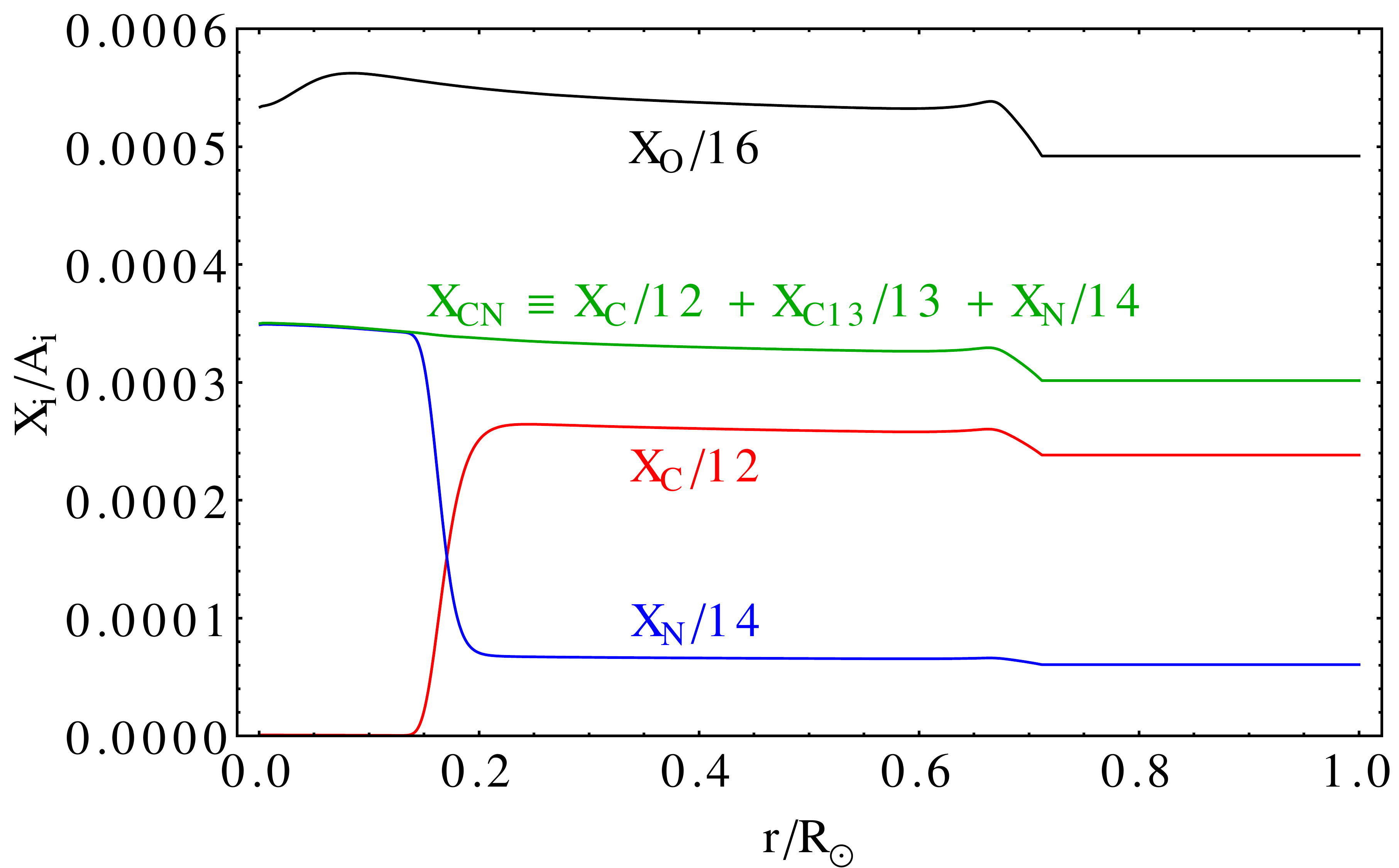}
\caption{\small\em \texttt{\rm Left Panel}: The behaviour of temperature $T$ and density $\rho$ (scaled to central values $T_{\rm c}$ and $\rho_{\rm c}$) and of mass $m$ and luminosity $l$ (scaled to total mass $M_\odot$ and luminosity $L_\odot$) as a function of the solar radius. 
%The central temperature is $T_{\rm c}=...$ ($...$) and the central density is $\rho_{\rm c}=...$ ($....$) in B16-HZ SSM (B16-LZ SSM);
\texttt{\rm Middle Panel}: The abundances of hydrogen ($X$), helium-4 ($Y$) and helium-3 ($X_3$) in the present Sun;
\texttt{\rm Right Panel}: The abundances of CNO elements in the present Sun.
\label{Fig:SSM}}
\end{center}
\end{figure}

The SSM predicts that most of the solar energy ($>99\%$) is produced by the pp-chain, i.e. the hydrogen fusion reaction chain displayed in the left panel of Fig.\ref{fig:PPandCNO}.
The pp-chain is is mostly initiated by $\ppS$ reaction and, to a minor extent, by electron capture reaction $\pepS$ and has several possible terminations that depend on the specific mechanism by which helium-3 nuclei, which are produced by $\DpS$ reaction, are converted to heavier elements. 
In the Sun, the dominant mechanism is 
$\hethetS$ that corresponds to the so-called pp-I termination of the pp-chain.
Alternatively, helium-3 can undergo $\hetheqS$ reaction with the effect of producing beryllium-7. 
Depending on the destiny of $^{7}{\rm Be}$, that can be processed either by the electron capture 
$\beeS$ or by the (largely sub-dominant) proton capture reaction $\bepS$, one obtains the pp-II or the pp-III terminations of the chain.
Finally, a very small amount of helium-4 nuclei is produced by $\hepS$ reaction.
The relative importance of the different branches of the pp-chain depends primarily on the core temperature of the Sun and on the cross section of specific reactions, as will be discussed in next section. 
The numbers given in Fig.\ref{fig:PPandCNO} show the branching ratios in the present Sun.
According to SSM calculations, the central temperature and density of the present Sun are $\tc\simeq 15.6 \times 10^6 \; {\rm K}$ and $\rhoc \simeq 150\; {\rm g \, cm^{-3}}$ and they decrease as a function of the solar radius as it is shown in Fig.\ref{Fig:SSM}.
Most of the solar luminosity is produced in the region $r \lsim 0.2\, R_\odot $ that contains about $30\%$ of the total mass of the Sun. 
In this region we observe a relevant increase (decrease) of the helium-4 (hydrogen) mass fraction $Y$ ($X$), as a result of hydrogen burning during the Sun lifetime.
The helium-3 mass fraction ($X_{3}$) has a non monotonic behaviour, explained by the fact that ${^{3}{\rm He}}$ burning time is larger than the age of the Sun for $r\gsim 0.3 R_\odot$ and thus helium-3 accumulates proportionally to the efficiency of $\DpS$ reaction. 
In the energy-producing core, however, $^{3}{\rm He}$ nuclei are efficiently converted to heavier elements by nuclear processes (mainly by $\hethetS$), and the abundance $X_{3}$ is equal to the equilibrium value.

An alternative hydrogen burning mechanism is provided by the CNO-bicycle that is displayed in the right panel of Fig.\ref{fig:PPandCNO}.
The CNO-bicycle uses carbon, nitrogen and oxygen nuclei that are present in the core of the Sun as catalysts for hydrogen fusion. 
It is composed by two different branches, i.e. the CN-cycle and the NO-cycle, whose relative importance depends on the outcome of proton capture reaction on nitrogen-15.
In the Sun, the 
%$^{15}{\rm N}+{\rm p} \to {^{12}{\rm C}} +\alpha$
$^{15}{\rm N}({\rm p},\alpha){^{12}{\rm C}}$
channel is largely dominant and so, in practice, the CNO-bicycle is reduced to the CN-cycle with a marginal contribution by the NO-cycle.  
Note that the CN-cycle conserves the total number of $^{12}$C and $^{14}$N nuclei in the core of the Sun, but alters their distribution as it burns into equilibrium, eventually achieving equilibrium abundances proportional to the inverse of the respective rates, see right panel of Fig.\ref{Fig:SSM}.
The reactions controlling conversion of $^{12}$C and $^{14}$N in the solar core and the approach  to equilibrium  are  $\CpS$ and  $\NpS$: these are  the next-to-slowest and  slowest rates in the CN-cycle, respectively.  
The temperature above which the $^{12}$C burning time through $\CpS$ is smaller than the Sun's lifetime is $T \sim 10^7 \,{\rm K}$.  In the SSM, the entire  energy-producing core, $r \lsim  0.2 R_\odot$ and $m \lsim 0.3 M_\odot$ is at temperature larger than this value, so that nearly  all of the core's carbon-12 is converted  to nitrogen-14. The slower $\NpS$  reaction determines whether  equilibrium is achieved.  
The $^{14}$N  burning time is shorter than the age of  the Sun for $T \gsim 1.3\times 10^{7}\,{\rm K}$.  Therefore equilibrium for the CN cycle is reached only for $R \lsim  0.1 R_\odot$, corresponding to the central 7\% of the  Sun by mass. Consequently, over a significant portion  of the outer core, $^{12}$C is converted   to  $^{14}$N,  but   further  reactions   are  inhibited   by  the $\NpS$ bottleneck. 

%------------------------------------------------------------
\begin{table}[t]
%\centering
%\begin{adjustwidth}{-0.5cm}{}
%\setlength{\tabcolsep}{1pt}
\footnotesize
\begin{center}
\begin{tabular}{c| c c }
\hline
\hline
Flux & B16-HZ & B16-LZ  \\
\hline
$\phipp$ & $5.98(1 \pm 0.006)$ & $6.03(1 \pm 0.005)$\\
$\phipep$  &$ 1.44(1 \pm 0.01) $&$1.46(1 \pm 0.009)$\\
$\phihep$ & $7.98(1 \pm 0.30) $&$8.25(1 \pm 0.30) $\\
$\phibe$ &$ 4.93(1 \pm 0.06)$ &$4.50(1 \pm 0.06) $ \\
$\phib$ & $5.46(1 \pm 0.12)$ &$4.50(1  \pm 0.12) $\\
$\phin$ & $2.78(1 \pm 0.15)$ &$2.04(1  \pm  0.14)$ \\
$\phio$ & $2.05(1 \pm 0.17)$ &$1.44(1 \pm 0.16) $\\
$\phif$ & $5.29(1 \pm 0.20)$ &$3.26(1 \pm 0.18) $\\ 
%\noalign{\smallskip}
\hline 
%\noalign{\smallskip}
$\phien$ & $2.20(1 \pm 0.15)$ &$1.61(1  \pm  0.14) $ \\
$\phieo$ & $0.81(1 \pm 0.17)$ &$0.57(1 \pm 0.16) $\\
$\phief$ & $3.11(1 \pm 0.20)$ &$1.91(1 \pm 0.18) $\\
%\noalign{\smallskip}
\hline 
%\noalign{\smallskip}
\end{tabular}
%\end{adjustwidth}
\end{center}
\caption{\small\em Solar neutrino fluxes predicted by SSMs with different surface composition \citep{Vinyoles:2017}. 
Units are: $10^{10}\,{\rm (pp)}$, 
$10^{9}\,{\rm(^7 Be)}$, 
$10^8\,\rm{(pep,\,^{13}N,\,^{15}O})$, $10^6\,(\rm{^8B,^{17}F)}$, 
$10^5\,\rm{(eN, eO)}$ 
and $10^3\,\rm{(hep, eF)}$ $\rm{cm^{-2} s^{-1}}$.}
\label{tab:neutrinos}
\end{table}
%--------------------------------------------------------------

\begin{figure}[!t]
\begin{center}
%\resizebox{0.44\textwidth}{!}{\includegraphics{figure3.eps}}
 \includegraphics[height=6.6cm]{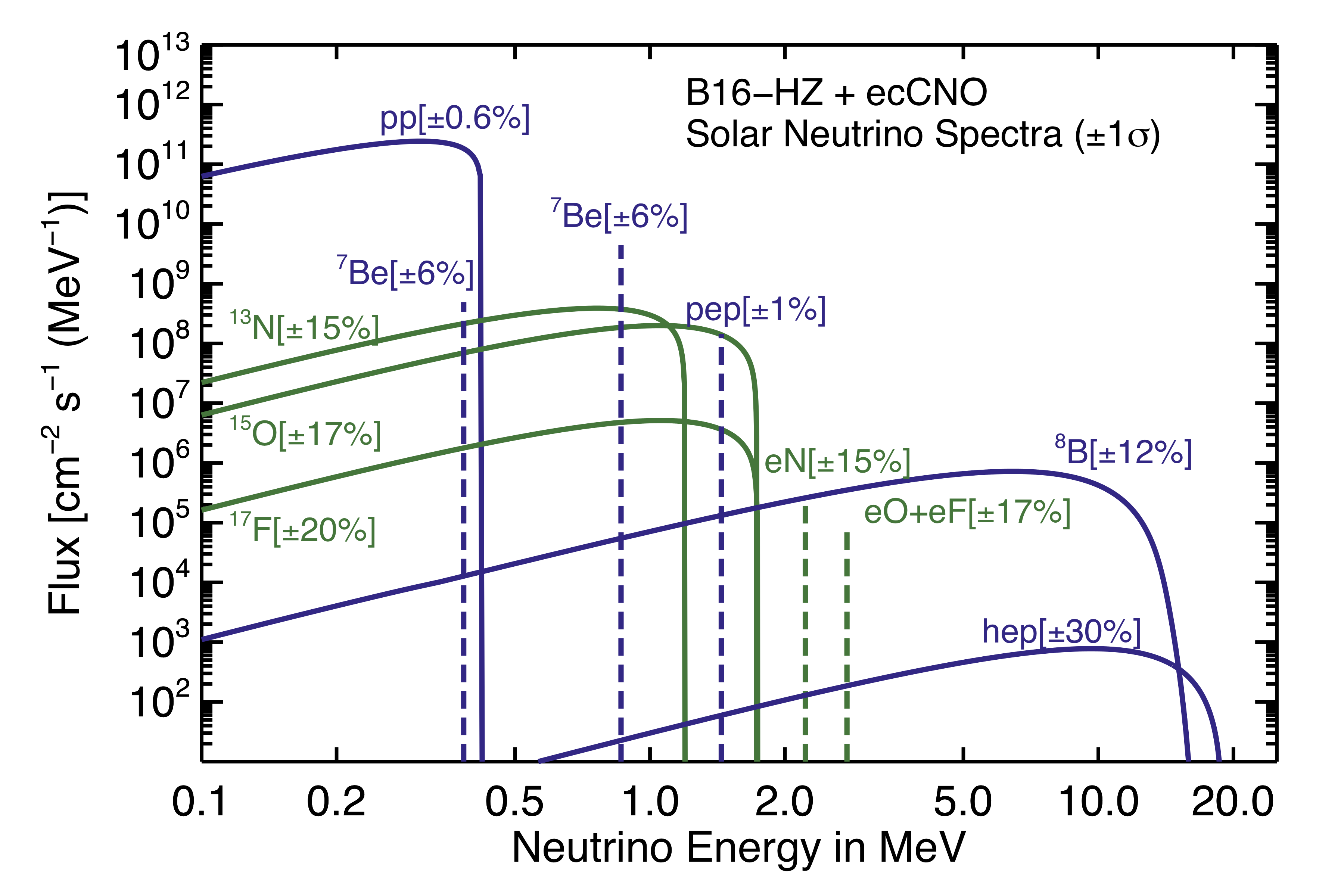}
%\resizebox{0.50\textwidth}{!}{\includegraphics{FiguraBorex.pdf}}
\caption{\small\em The solar neutrino spectrum.
\label{fig:nuspec}}
\end{center}
\end{figure}

A very effective tool to investigate nuclear energy generation in the Sun is provided by neutrinos which are necessarily produced along with $^{4}{\rm He}$ nuclei during hydrogen burning, in order to satisfy lepton number conservation. 
Neutrinos free stream in the solar plasma and reach the Earth in about $8$ minutes where they can be detected by solar neutrino experiments. 
While the total amount of neutrinos produced in the Sun can be easily estimated from the solar luminosity constraint, i.e. the assumption that the luminosity radiated from the surface of the Sun is exactly counterbalanced by the amount of energy produced by hydrogen fusion reactions in the solar core (see e.g. \cite{Bahcall:2001,DeglInnocenti:1997,Vissani:2018} for a detailed discussion), the evaluation of their spectrum requires the knowledge of the individual rates of neutrino producing reactions and thus the construction of a complete solar model. 
We report in Fig.\ref{fig:nuspec} and Tab.\ref{tab:neutrinos}, the SSM predictions for the different components of the solar neutrino flux, named according to the specific reaction by which they are produced \citep{Vinyoles:2017}.
We also include, for completeness, ecCNO neutrinos, i.e. neutrinos produced by electron capture reaction in the CNO-bicycle (in addition to the "standard" CNO neutrinos produced by $\beta$ decays of $^{13}{\rm N}$, $^{15}{\rm O}$ and $^{17}{\rm F}$) that were originally calculated in \cite{Bahcall:1990} and \cite{Stonehill:2004} and recently reevaluated in \cite{Villante:2015}\footnote{In order to take into account the new inputs in B16-SSM calculations, the ecCNO fluxes given in Tab.\ref{tab:neutrinos} have been scaled with respect to the values quoted  in \cite{Villante:2015} proportionally to the corresponding $\beta$-decay fluxes. This follows from the assumption that the ratio of electron capture and beta decay processes in the Sun is equal to what evaluated in \cite{Villante:2015}}. 
The two columns "B16-HZ" and "B16-LZ" reported in Tab.\ref{tab:neutrinos} are obtained by considering two different options for the solar surface composition, as it discussed in Sect.\ref{subsec:compo}.
During the last few decades, solar neutrino experiments have allowed us to determine with great accuracy most of the components of the solar flux. 
As an example, $^{7}{\rm Be}$ and $^{8}{\rm B}$ neutrino fluxes are measured with accuracy better than $\sim 3\%$ by Borexino \citep{Borexino:2018}, Super-Kamiokande \citep{SK:2016} and SNO \citep{SNO:2013}.
The pp and pep-neutrino flux can be determined with $\lsim 1\%$ accuracy by assuming the solar luminosity constraint, see e.g.~\cite{Bergstrom:2016}.
These fluxes, however, have been also directly measured by Borexino \citep{Borexino:2012,Borexino:2014,Borexino:2018} with $\sim 10\%$ and $\sim 17\%$ accuracy, respectively.
Finally, Borexino has recently obtained the experimental identification of CNO neutrinos \citep{Borexino:CNO}, providing the first direct evidence that CNO-bicycle is active in the Sun. 

\vspace{0.5cm}

\subsection{Nuclear reaction rates}

The cross sections of nuclear reaction in pp-chain and in CNO-bicycle are fundamental inputs for SSM calculations. 
Even if the focus of this work is on the role of nuclear rates for solar modeling (more than on reviewing the present situation for cross section measurements and calculations), we believe that it is useful to briefly discuss the adopted assumptions for the B16-SSM~\citep{Vinyoles:2017}. 
%that are taken as a reference and 
whose results have been previously discussed. 
The nuclear rates adopted for these models are from the Solar Fusion II compilation \citep{Adelberger:2011} with few relevant changes summarized in the following.

\begin{itemize}
 
\item $\boldsymbol{{\rm p(p,e^+\nu_e)d}}$: The astrophysical factor $S_{11}(E)$ has been recalculated in \cite{Marcucci:2013} by using  chiral effective field theory framework, including the P-wave contribution that had been previously neglected. 
For the leading order they obtain $S_{11}(0)= (4.03 \pm 0.006) \cdot 10^{-25}\,\rm{MeV\,b}$. 
More recently, and also using chiral effective field theory, $S_{11}(E)$ was calculated by \cite{Acharya:2016}, resulting in $S_{11}(0)= 4.047^{+0.024}_{-0.032}\cdot 10^{-25}\,\rm{MeV\,b}$. This is in very good agreement with result from \cite{Marcucci:2013}. \cite{Acharya:2016} have performed a more thorough assessment of uncertainty sources leading to an estimated error of 0.7\%, much closer to the 1\% uncertainty which was obtained by \cite{Adelberger:2011}. 
In B16-SSM calculations, the astrophysical factor $S_{11}(E)$ is taken  from \cite{Marcucci:2013} with a conservative 1\% error estimate \citep{Vinyoles:2017}.

\item $\boldsymbol{{\rm ^7Be(p,\gamma)^8B}}$: Solar Fusion II recommended value is $S_{17}(0) = (2.08 \pm 0.07 \pm 0.14)\cdot 10^{-5} \,\rm{MeV\,b}$ \citep{Adelberger:2011}, where the first error term comes from uncertainties in the different experimental results and the second one from considering different theoretical models employed for the low-energy extrapolation of the rate. \cite{Zhang:2015} presented a new low-energy extrapolation $S_{17}(0) = (2.13 \pm 0.07)\cdot 10^{-5} \,\rm{MeV\,b}$, based on Halo Effective Field Theory, which allows for a continuous parametric evaluation of all low-energy models. Marginalization over the family of continuous parameters then amounts to marginalizing the results over the different low-energy models.
In B16-SSM calculations, it was conservatively adopted an intermediate error between those from \cite{Zhang:2015} and \cite{Adelberger:2011}. The adopted value is $S_{17}(0) = (2.13 \pm 0.1)\cdot 10^{-5} \,\rm{MeV\,b}$. The derivatives of the astrophysical factor were updated by using the recommended values in \cite{Zhang:2015}.

\item $\boldsymbol{{\rm ^{14}N(p,\gamma)^{15}O}}$: \cite{Marta:2011} presented cross-section data for this reaction obtained at the Laboratory for Underground Nuclear Astrophysics (LUNA) experiment. With the new data and using R-matrix analysis they recommend the value for the ground-state capture of $S_{GS}(0) = (0.20 \pm 0.05 ) \cdot 10^{-3} \hspace{1mm}\rm{MeV\,b}$. Combined with other transitions (see Table XI in that work) this leads to  $S_{114}(0) = (1.59 \cdot 10^{-3}) \hspace{1mm}\rm{MeV\,b}$, about 4\% lower than the previous recommended value in~\cite{Adelberger:2011}. The derivatives and the errors remain unchanged.

\item $\boldsymbol{{\rm ^3He(^4He,\gamma)^7Be}}$: 
Two recent analyses \citep{deBoer:2014,Iliadis:2016} have provided determinations of the astrophysical factor that differs by about $6\%$ (to be compared with a claimed accuracy equal to $4\%$ and $2\%$ for \cite{deBoer:2014} and \cite{Iliadis:2016}, respectively). Considering that the results from \cite{deBoer:2014} and \cite{Iliadis:2016} bracket the previously adopted value from \cite{Adelberger:2011}, the latter was considered as preferred choice in \cite{Vinyoles:2017}.

\end{itemize}

Finally, Salpeter's formulation of weak screening \citep{Salpeter:1954} is adopted. The validity of this formulation for solar conditions, where electrons are only weakly degenerate, has been discussed in detail in \cite{Gruzinov:1998}, where a more sophisticated approach was shown to lead, to within differences of about 1\%, to Salpeter's result. Other proposed deviations from this formulation have been discussed at length in \cite{Bahcall:2002}, including different approaches to dynamic screening, and shown to be flawed or not well physically motivated. More recent calculations of dynamic screening \citep{Mao:2009,Mussack:2011} still leave, however, some room for discussion on this topic. In the weak screening limit, and in conditions under which screening is not numerically large, the dominant scaling is with the product of the charge of the two reacting nuclei. In the solar core, screening enhancement is about 5\% for $\ppS$, 20\% for $\hetheqS$ and $\bepS$, and 40\% for $\NpS$.

\section{The role of nuclear reactions}
\label{Sec:role}
In the following, we discuss the role of nuclear reactions in SSM construction. Among nuclear processes, the $\ppS$ reaction is the only one that can affect the temperature stratification of the Sun.
Indeed, this process determines the global efficiency of hydrogen burning in the Sun. 
The other reactions in the pp-chain and in the CNO-cycle have a minor importance in this respect. 
However, they have a crucial role in determining the relative rates of the different pp-chain terminations and the efficiency of the CNO-cycle, thus affecting the predictions for the different components of the solar neutrino spectrum.

\vspace{0.5cm}

\subsection{The pp-reaction rate and the central temperature of the Sun}
\label{subsec:ppandtc}
In SSM calculations, where the Sun is assumed to be in thermal equilibrium, the rate of the pp-reaction is basically determined by the solar luminosity.
Indeed, by considering that helium-4 is mainly produced by $\hethetS$, we arrive at the conclusion that the integrated pp-rate in the Sun is $\Rpp \sim 2 L_\odot/Q_{\rm I}$, where $Q_{\rm I} = Q - 2 \langle E_\nu \rangle_{\rm pp} \sim 26.2 \,{\rm MeV}$ is the energy released in the solar plasma when $^{4}{\rm He}$ is synthesized through pp-I termination.
In the previous expression, we considered that the average energy of neutrinos produced by $\ppS$ is $\langle E_\nu \rangle_{\rm pp} = 0.265 \,{\rm MeV}$
and we took into account that, at equilibrium, the pp-I termination involves twice the pp-reaction in order to feed the process $\hethetS$. 

\begin{figure}[t]
\begin{center}
\includegraphics[width=8.7cm]{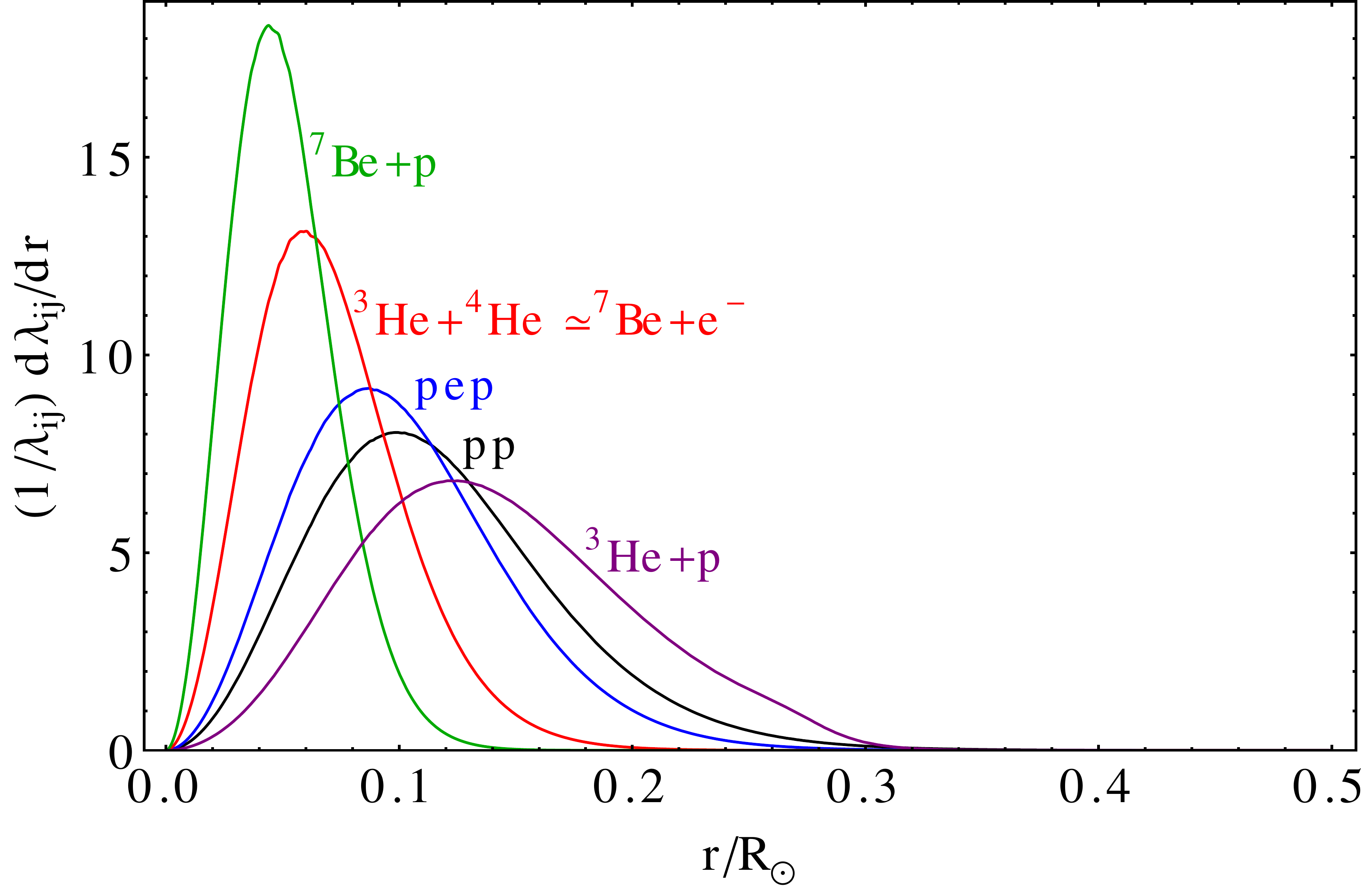}
\includegraphics[width=8.7cm]{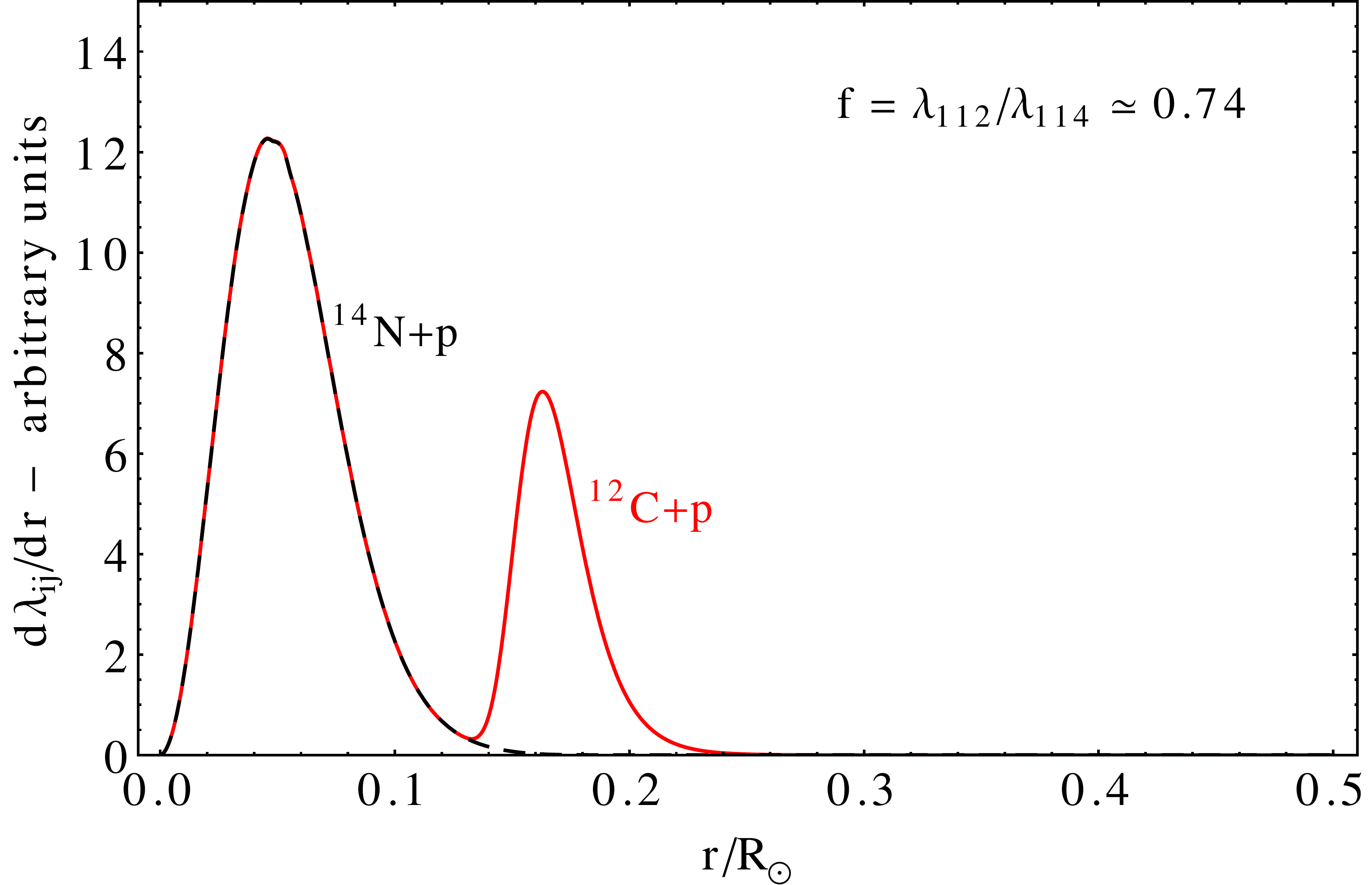}
\caption{\small\em 
The differential rates for nuclear reactions in the pp-chain (\texttt{\rm Left Panel}) and CN-cycle (\texttt{\rm Right Panel}).
The curves in the left panel have been normalized to one to facilitate comparison among them.  
The curves in the right panel are not normalized to emphasize that reactions $\CpS$ and $\NpS$ have the same rate in the equilibrium region.
\label{Fig:Rates}}
\end{center}
\end{figure}

Being the reaction rate fixed by the observed luminosity, the cross section of $\ppS$ determines the central temperature of the Sun, as it is explained in the following. 
The rate $\Rpp$ can be expressed as:
\begin{equation}
\Rpp= \int d^3r \; \frac{\rho^2}{m^2_{\rm
    u}} \, \frac{X^2}{2} \, \svpp
\end{equation}
where $\rho$ is the density, $m_{\rm u}$ is the atomic mass unit, $X$ is the hydrogen mass fraction and $\svpp$ is the reaction rate per particle pair of the $\ppS$ reaction. 
The above integral involves, in principle, the entire solar structure but it gets a non-vanishing contribution only from the inner core of the Sun at $r \le 0.3 \,R_{\odot}$. 
This can be appreciated by looking at  Fig.~\ref{Fig:Rates} where we show the differential rates $(1/\lambda_{\rm ij})\,(d\lambda_{\rm ij}/dr)$ for the $\ppS$ (black), $\pepS$ (blue), $\hetheqS$ (red), $\bepS$ (green) and $\hepS$ (purple) reactions as a function of the solar radius.
The different curves are all normalized to one in order to facilitate comparison among them.
These curves also corresponds to the normalized production rates of pp, pep, $^7$Be, $^8$B and hep neutrinos, respectively\footnote{Note that the rate of $\hetheqS$ is basically equal to that of the neutrino producing reaction $\beeS$}. 

\begin{figure}[t]
\begin{center}
\includegraphics[width=8.7cm]{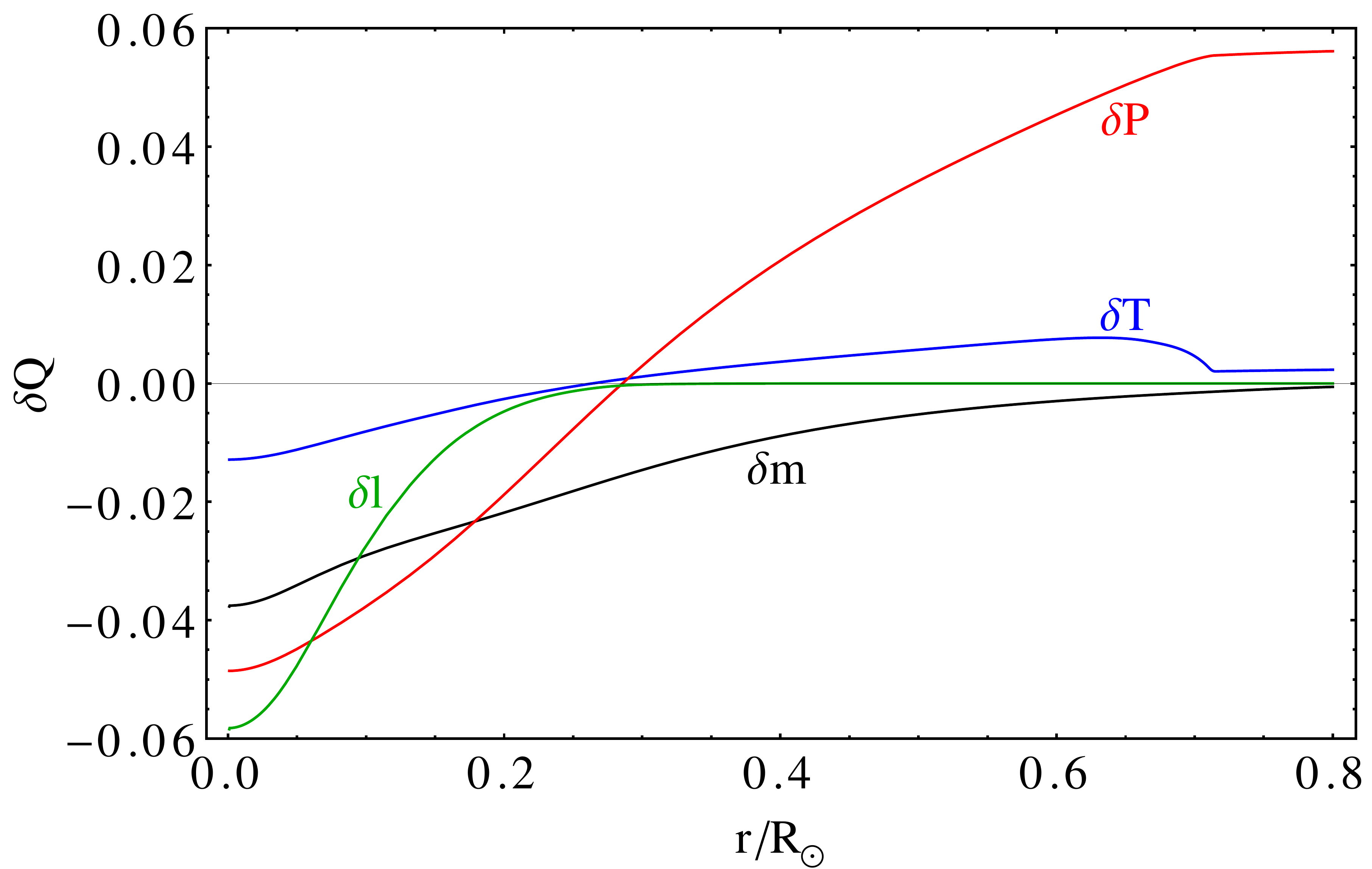}
\includegraphics[width=8.7cm]{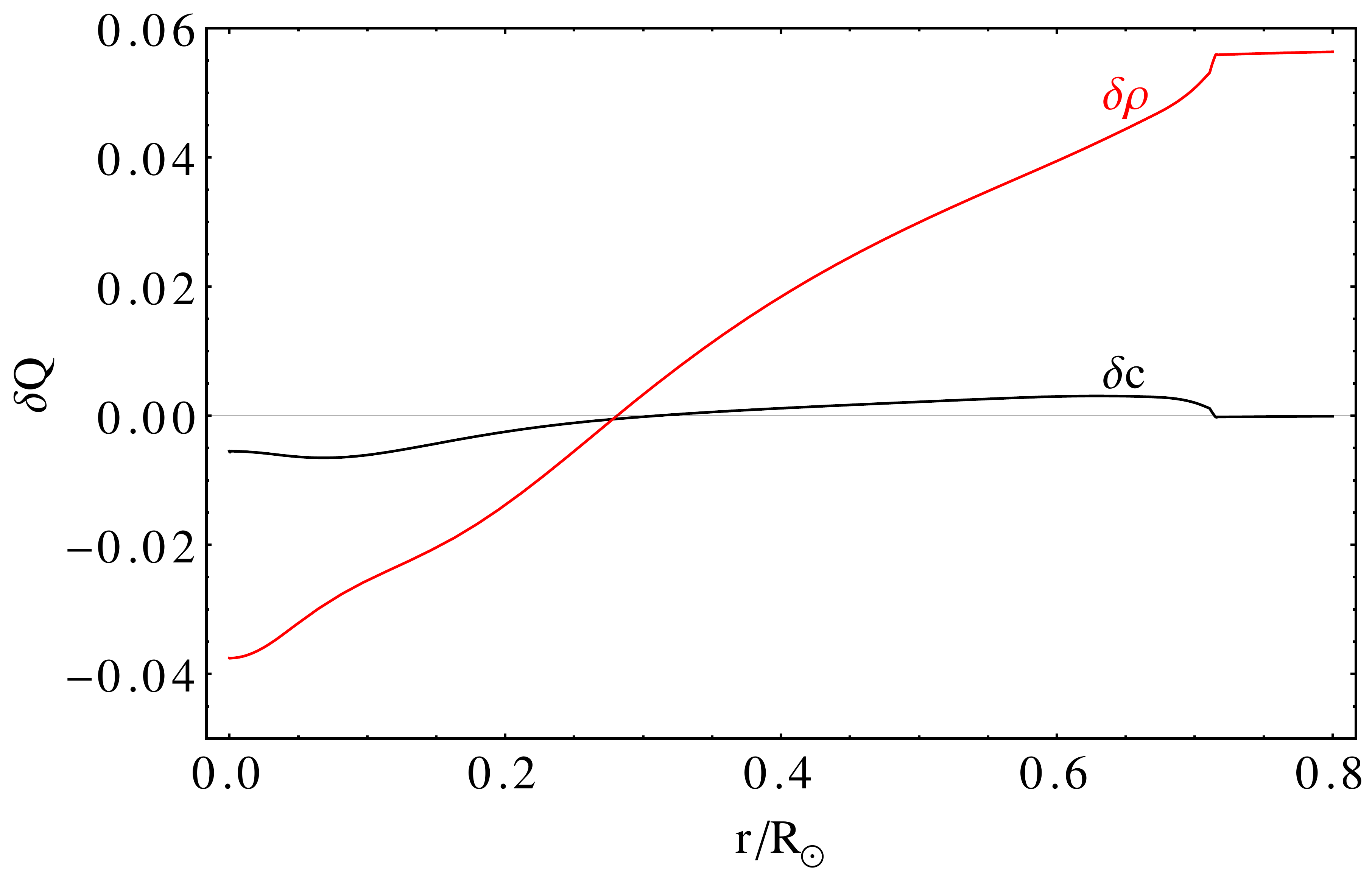}
\caption{\small\em The effects of a $10\%$ increase of the astropysical factor of $\ppS$ reaction on the physical properties of the Sun (\texttt{\rm left Panel}) and on helioseismic observable quantities $\delta c(r)$ and $\delta \rho(r)$ (\texttt{\rm right Panel}). The fractional variations $\delta Q$ are calculated with respect to the reference SSM predictions.
\label{Fig:Spp}}
\end{center}
\end{figure}

Taking into account that $\ppS$ reaction is active in a narrow region of the Sun at $r_0\simeq 0.1 R_\odot$ whose physical conditions are similar to those at the solar center, we write the approximate scaling law: 
\begin{equation}
\Rpp \propto  \rhoc^2 \, \xcen^2 \, \Spp \, \tc^{\gammapp}
\label{eq:rpp-propto}
\end{equation}
where the notation $\Qc$ indicates that the generic quantity $Q$ is evaluated at the center of the Sun, $\Spp$ is the astrophysical factor of the pp-reaction and  we considered that $\svpp \propto \Spp \, \tc^{\gammapp}$ with $\gammapp \simeq 4$.
%
%\footnote{The exponent $\gamma_{ij}$ can be estimated as $\gamma_{ij}=(E_{0})_{ij}-2/3$ where $(E_{0})_{ij}$ is the Gamov peak energy of the considered reaction}
%
Eq.(\ref{eq:rpp-propto}) implies the following linearised relationship:
\begin{equation}
\nonumber
\delta \Rpp \simeq  2 \delta \rhoc + 2 \delta \xcen +\gammapp \, \delta \tc +\delta  \Spp \; ,
\label{eq:rpp-linear}
\end{equation}
where $\delta Q$ indicates the fractional variation of the quantity $Q$ with respect to the reference SSM value.
The above expression contains input parameters for solar model construction, i.e. the astrophysical factor $\Spp$, and structural parameters, like e.g. the temperature, density and hydrogen  abundance in the core of the Sun which are the result of solar model self-calibrated calculations. In principle, a modification of $\Spp$ induces a change of the solar structure and, thus, the different terms in the r.h.s of eq.(\ref{eq:rpp-linear}) are correlated. 
In order to keep $\delta \Rpp \simeq 0$, an increase of the astrophysical factor $\delta \Spp \ge 0$ has to be counterbalanced by an opposite contribution $2 \delta \rhoc + 2 \delta \xcen +\gammapp \, \delta \tc \le 0$. This is achieved by varying the initial helium and metal abundance of the Sun according to $\delta \yini = 0.6\, \delta\Spp $ and $\delta \zini \simeq -0.10 \,\delta \Spp $ with the effect 
of obtaining a (slightly) colder solar core. We obtain numerically:
\begin{equation}
\delta \tc \sim -0.13 \, \delta \Spp
\label{eq:deltatc-linear}
\end{equation}
that will be useful in the following to understand the effects of $\Spp$ variations on the various components of the solar neutrino spectrum. 
In Fig.\ref{Fig:Spp}, we show the effect of a 10\% increase of $\Spp$ on the temperature profile of SSMs and on the helioseismic observable quantities $\delta c(r)$ and $\delta \rho(r)$.

\vspace{0.5cm}

\subsection{The dependence of neutrino fluxes on the central temperature of the Sun and on nuclear reaction cross sections}
\label{subsec:neutrinos}
Even a small modification of the central temperature of the Sun reflects into large variations of solar neutrino fluxes.
By considering the arguments discussed in \cite{Bahcall:1996} and \cite{DeglInnocenti:1997}, we discuss the dependence of solar neutrino fluxes on the core temperature of the Sun, highlighting the role of nuclear reactions for determining the branching ratios of the different pp-chain terminations and the efficiency of the CNO-bicycle.

\vspace{0.5cm}

\subsubsection{PP-chain neutrino fluxes}
\label{susec:ppchain}

\noindent {\em The pp-neutrino flux:} \\
The vast majority of the solar neutrino emission is due to pp-neutrinos whose flux $\phipp$ is directly linked to $\Rpp$ being $\phipp = \Rpp/(4 \pi D^2)$ where $D=1\, {\rm A.U.}$ is the Sun-Earth distance.
According to discussion in the previous Section, the rate $\Rpp$ is directly fixed by solar luminosity and thus $\phipp$ is expected to be independent from the central temperature of the Sun and nuclear reaction cross sections.
This result is obtained by assuming that pp-I is the only  mechanism for helium-4 production by nuclear reaction in the Sun. 
A more accurate description can be obtained by taking into account the contribution the secondary branches of the pp-chain (namely, the pp-II termination) initiated by the $\hetheqS$ which provides an alternative $^{3}{\rm He}$ burning mechanism to the most common $\hethetS$.
In this assumption, we have:
\begin{equation}
L_{\odot} = Q_{\rm I} \, \Rtt + Q_{\rm II} \, \Rtq
\end{equation}
where $\Rtt$ and $\Rtq$ are the integrated rate of the $\hethetS$ and $\hetheqS$ reactions, while $Q_{\rm I}= 26.20\,{\rm MeV}$ and $Q_{\rm II} = 25.65\,{\rm MeV}$ give the amount of energy, corrected for neutrino emission, delivered in the plasma when $^4{\rm He}$ is produced through pp-I and pp-II terminations, respectively. 
By considering that $\Rpp = 2 \Rtt + \Rtq$ at equilibrium, we arrive at the conclusion that \citep{Bahcall:1996}:
\begin{equation}
\phipp = \frac{1}{4\pi D^2}\left(\frac{2\lsun}{Q_{\rm I}} - \Rtq\right)
\end{equation}
where we considered that $Q_{\rm I}\simeq Q_{\rm II}$. 
While the first term in the r.h.s. of the above equation is constant, the rate $\Rtq$ depends on the temperature of the plasma and on nuclear reaction cross sections. If we take into account that $\Rtq\propto S_{34}\cdot (S_{11} / S_{\rm 33})^{1/2} \cdot \tc ^{\beta_{\rm Be}}$ with $\beta_{\rm Be} \sim 11$, as motivated later in this section, we obtain the following  relationship
\begin{equation}
\delta \phipp = -\eta \, \delta\Stq -  \frac{\eta}{2}\left(\delta \Spp - \delta \Stt \right) + 
\beta_{\rm pp} \, \delta \tc
\label{eq:deltaphipp}
\end{equation}
that gives the fractional variation of the flux $\delta\phipp$ as a function of fractional variation of the core temperature $\delta \tc$ and of the astrophysical factors $\delta \Sij$.
The coefficients in the above equation correspond to the logarithmic derivatives of $\phipp$ with respect to these quantities and are given by $\eta = \Rtq/\Rpp \simeq \phibe/\phipp \simeq 0.08$ and $\beta_{\rm pp}= - \eta \beta_{\rm  Be} \simeq -0.9$, showing that the pp-neutrino flux is a decreasing function of the central temperature of the Sun.

\noindent{\em The pep-neutrino flux:}\\
The pep-neutrinos are produced by electron capture reaction $\pepS$ which is linked to the $\beta$-decay process $\ppS$ by well-known nuclear physics. Since the two processes depend on the same allowed nuclear matrix element, the ratio between their rates is determined by the available reaction phase spaces and by the electron density $n_e$ of the solar plasma only. 
It can be determined with $\sim 1\%$ precision for the conditions of the solar interior and is mildly dependent on the properties
of the solar plasma, being roughly proportional to $\tc^{-1/2}\, n_{e}$ (see e.g. \cite{Adelberger:2011} for a review). We can thus assume $\phipep\propto \tc^{1/2}\, \phipp$,
%\begin{equation}
%\phipep \propto S_{34}^{-a}\cdot (S_{11} / S_{\rm %33})^{-a/2} \cdot \tc ^{\beta_{\rm pep}}
%\end{equation}
allowing us to conclude: 
\begin{equation}
\delta \phipep  = 
-\eta \, \delta\Stq -  \frac{\eta}{2}\left(\delta \Spp - \delta \Stt \right) + 
\beta_{\rm pep} \, \delta \tc
\label{eq:deltaphipep}
\end{equation}
where $\beta_{\rm pep} = \beta_{\rm pp}-1/2 \simeq -1.4$, and we neglected effects related to possible density and chemical composition variations in the solar core.

\noindent{\em The $^7$Be-neutrino flux:}\\
The formation of beryllium-7 through $\hetheqS$ leads to neutrino production through the electron capture reaction $\beeS$. 
This process largely dominates over the  competing proton-capture reaction whose effects are discussed in the following paragraph. 
Taking this into account, the Be-neutrino flux can be directly estimated from the rate 
of the $\hetheqS$ reaction
by using $\phibe = \Rtq /(4\pi D^2)$.
The rate $\Rtq$ is given by:
\begin{equation}
\Rtq = 
\int d^3r \; 
\frac{\rho^2}{m^2_{\rm    u}} 
\, \frac{X_{3}\,Y}{12} \, 
\svtq
\label{eq:r34}
\end{equation}
where $Y$ ($X_{3}$) is the helium-4 (helium-3) mass fraction and $\langle \sigma v \rangle_{34}$ is the reaction rate per particle pair of $\hetheqS$.
The amount of helium-4 nuclei in the present Sun is determined by the assumed initial abundance $Y_{\rm ini}$ and by nuclear processes that have converted hydrogen into helium during the Sun evolution. 
We may thus expect that $Y$ depends on nuclear cross sections, in particular on $\Spp$ that determines the global efficiency of hydrogen burning.
This dependence is however marginal
because the product $\lsun\tausun$ essentially provides an observational determination of the  
integrated solar luminosity (and thus of the total amount of helium synthesized by nuclear reactions during the Sun lifetime).
The helium-3 abundance in the solar core depends instead on the temperature $\tc$ and on the cross sections of the $\ppS$ and $\hethetS$ reactions.
It can be indeed calculated by using the equilibrium condition 
\begin{equation}
X_3 \simeq X_{3, \rm eq} = 3\, X \sqrt{\frac{\svpp}{2 \svtt}}
%\right)^{1/2}
\end{equation}
where $X$ is the hydrogen mass fraction.
Considering that $\svij \propto \Sij \; \tc ^{\gammaij}$, this 
can be rewitten as $X_{3,\rm c} \propto (S_{11} / S_{\rm 33})^{1/2} \cdot \tc ^{(\gamma_{11}-\gamma_{33})/2}$ where we neglected effects related to possible  hydrogen abundance variations\footnote{We evaluate the exponents $\gamma_{ij}$ by using $\gamma_{ij}\simeq (E_{0})_{ij}-2/3$ where $(E_{0})_{ij}$ is the Gamov peak energy of the considered reaction, see e.g. \cite{Bahcall:1996}}. 
This expression, combined with Eq.~(\ref{eq:r34}), allows us to conclude that:
\begin{equation}
\phibe 
%\simeq
%frac{\Rtq}{4 \pi D^2}
\propto
S_{\rm 34} \cdot (S_{11} / S_{\rm 33})^{1/2} \cdot \tc ^{\beta_{\rm Be}}
\end{equation}
or, equivalently,
\begin{equation}
\delta \phibe 
=
\delta \Stq 
+\frac{1}{2} 
\left(\delta \Spp - \delta \Stt \right) 
+ \beta_{\rm Be} \, \delta \tc
\label{eq:deltaphibe}
\end{equation}
where $\beta_{\rm Be} = \gammatq +(\gammapp-\gammatt)/2 \sim 11$.
Note that the $^7$Be-neutrino flux does not depends on the cross section of $\beeS$,
due to the fact that (almost) the totality of beryllium-7 nuclei produced by $\hetheqS$ are expected to decay through this reaction.

\noindent{\em The $^8$B-neutrino flux:}\\
The $^8$B neutrinos constitute a largely subdominant component of the solar flux which is produced when $^{7}{\rm Be}$ nuclei capture a proton (instead of an electron) producing  $^{8}{\rm B}$ (instead of $^{7}{\rm Li}$). 
The $^8$B-neutrino flux is thus given by $\phib = r \, \phibe$ where $r\equiv \Rbep/\Rbee$ is the ratio between proton and electron capture rates on beryllium-7. 
The parameter $r$ scales as $r \propto (\Sbep/\Sbee)\cdot \tc^{\alpha}$ where $\alpha=
\gammabep + (1/2)$
and we have considered that $\svbee \propto \Sbee \; \tc ^{-1/2}$ for electron capture reaction. 
Taking this into account, we obtain the following scaling law:  
\begin{equation}
\phib \propto \left( S_{17}/ S_{e7}\right) \cdot S_{\rm 34} \cdot \left(S_{11} / S_{\rm 33}\right)^{1/2} \cdot \tc ^{\beta_{\rm B}}
\end{equation}
that also corresponds to:
\begin{equation}
\delta \phib
=
\left(\delta \Sbep - \delta \Sbee \right)
+ \delta \Stq 
+\frac{1}{2} 
\left(\delta \Spp - \delta \Stt \right) 
+ \beta_{\rm B} \, \delta \tc
\label{eq:deltaphib}
\end{equation}
with $\beta_{\rm B} = \beta_{\rm Be} +\gammabep + 1/2 \simeq 24$. The large value of $\beta_{\rm B}$ indicates that $^8$B neutrinos are a very sensitive probe of the core temperature of the Sun. 

\vspace{1cm}

\subsubsection{The CNO neutrino fluxes}
The neutrino fluxes produced in the CN-cycle by $\beta$-decay (and electron capture reactions) of $^{13}$N and $^{15}$O nuclei, besides depending on the solar central temperature, 
%and on nuclear reaction cross sections, 
are approximately proportional to the stellar-core number abundance of CN elements. 
This dependence is relevant to understand the role of cross section for CNO-neutrino production.
Moreover, as it is discussed in \cite{Haxton:2008,Haxton:2013,Serenelli:2013}, it permits us to use CNO neutrinos, in combination with other neutrino fluxes,
%to break the composition-opacity degeneracy and, in principle, 
to directly probe the chemical composition of the Sun. 

\noindent{\em The $^{15}$O-neutrino flux:}\\
This component of the solar neutrino spectrum is determined by the production rate of oxygen-15 by $\NpS$ reaction in the core of the Sun. It can be calculated as $\phio = \RNp/(4\pi D^2)$ where
the rate $\RNp$, given by:
\begin{equation}
\lambda_{114} =  
\int d^3r \; \frac{\rho^2}{m^2_{\rm
    u}} \, \frac{X \,
  X_{\rm 14}}{14} \, 
  \svNp
  \label{eq:RNp}
\end{equation}
is proportional to the nitrogen-14 mass fraction $X_{14}$ in the solar core (see Fig.\ref{Fig:SSM}) and to the reaction rate per particle pair $\langle\sigma v\rangle_{114}$ of the $^{14}$N(p,$\gamma$)$^{15}$O reaction.
The above integral get a non vanishing contribution from a narrow region at $r \lsim 0.1 R_{\odot}$ whose conditions are similar to that at the solar center, see Fig.\ref{Fig:Rates}. We thus write the approximate scaling law: 
\begin{equation}
\phio \propto  
\RNp
\propto
X_{\rm 14,\rm c} \; \SNp \, \tc^{\beta_{\rm O}}
\label{eq:phio-propto}
\end{equation}
where $S_{114}$ is the astrophysical factor of the $^{14}$N(p,$\gamma$)$^{15}$O reaction, we considered that $\langle\sigma v\rangle_{114} \propto S_{114} \, \tc^{\gamma_{114}}$ and we defined $\beta_{\rm O} = \gamma_{114} \simeq 20$.
Eq.~(\ref{eq:phio-propto}) implies the following linearised relationship:
\begin{equation}
\delta \phio \simeq \delta X_{\rm 14,\rm c} +\delta \SNp + \beta_{\rm O} \; \delta \tc
\label{eq:phio-linear}
\end{equation}
In the above expressions, we neglected effect related to possible variations of the density and of the hydrogen abundance in the solar core, since these are expected to be small.
We instead explicitly considered the dependence of $\phio$ on the central abundance of nitrogen-14 which is essentially determined, as it is explained in the following, by the total abundances of CN elements in the solar core. 
It is useful to remark that, being the CNO cycle sub-dominant, a modification of its efficiency does not alter the solar luminosity and does not require a readjustement of the central temperature.
Moreover, carbon and nitrogen give a marginal contribution to the opacity of the solar plasma and thus a variation of their abundances do not alter the temperature stratification. 
As a result of this, we can consider the different terms in eq.(\ref{eq:phio-linear}) as independent.

\noindent{\em The $^{13}$N-neutrino flux:}\\
The flux of $^{13}$N-neutrinos can be calculated $\phin = \RCp/(4\pi D^2)$ where $\RCp$ is the total rate of the $\CpS$ reaction in the Sun. This is given by:
\begin{equation}
\RCp = 
\int d^3r \;
\frac{\rho^2}{m^2_{\rm u}} \, 
\frac{X \,  X_{\rm 12}}{12} \,
\svCp
%n_{\rm 1} \,  n_{\rm 12} \,
%\langle\sigma
%v\rangle_{112}
\end{equation}
where $X_{\rm 12}$ is the carbon-12 mass fraction and $\langle\sigma v\rangle_{112}$ is the reaction rate per particle pair of $^{12}$C(p,$\gamma$)$^{13}$N. We can write:
\begin{equation}
\RCp = \RNp + \RCp^{\rm (ne)}
\label{eq:RCpne-def}
\end{equation}
where the quantity:
\begin{equation}
\RCp^{\rm (ne)} =
\int d^3r \;
\frac{\rho^2}{m^2_{\rm u}} \, X \,
\left[\frac{X_{\rm 12}}{12} \, \svCp
     -\frac{X_{\rm 14}}{14}\, \svNp 
     \right]
\label{eq:ne-rate}
\end{equation}
gives the contribution to the total rate produced in the region of the Sun where the CN-cycle is incomplete.
The above integral vanishes indeed for $r \le 0.13 R_\odot$ where the equilibrium condition for the CN-cycle ensures that  $(X_{12}/12) \,\svCp - (X_{14}/14)\, \svNp =0$.
This can be appreciated in the right panel of Fig.\ref{Fig:Rates} where we show the differential rate $d\RNp/dr$ and $d\RCp/dr$ of $\NpS$ (black) and $\CpS$ (red) reactions as a function of the solar radius $r$.

Eq.(\ref{eq:RCpne-def}) implies that $\phin$ can be decomposed as the sum:
\begin{equation}
\phin = \phio + \phin^{\rm (ne)}
\label{eq:phio-ne-def}
\end{equation}
where the quantity $\phin^{\rm (ne)} \equiv \RCp^{\rm (ne)}/(4 \pi D^2)$ represents the neutrino flux produced in the region $0.13\lsim r/R_\odot \lsim 0.25$, where $\NpS$ reaction is not effective.
This component of the flux scales as:
\begin{equation}
\phin^{\rm (ne)} \propto \carboneq 
\, S_{112} 
\, \tc^{\gammaNp}
\label{eq:phin-ne-propto}
\end{equation}
where we considered that $\svCp \propto \SCp \, \tc^{\gammaCp}$ with $\gammaCp \simeq 18$
and we neglected effects related to possible variations of density and hydrogen abundance.
Note that the carbon-12 mass fraction in Eq.~(\ref{eq:phin-ne-propto}) is evaluated at $\rneq \simeq 0.16 \,R_\odot$ where the out-of-equilibrium $^{13}$N-neutrino production rate is maximal, see Figs.\ref{Fig:SSM} and \ref{Fig:Rates}.
In principle, the temperature should be also evaluated at this position. However, we can take the central value $\tc$ as representative for the entire energy producing region, motivated by the fact that $T(r)$ (differently from $X_{12}(r)$) is slowly varying in the solar core.
Eq.(\ref{eq:phin-ne-propto}) implies the following relationship: 
\begin{equation}
\label{eq:phin-ne-linear}
\delta \phin^{\rm (ne)} =\delta \carboneq +\delta \SCp + \gammaCp \; \delta \tc 
\end{equation}
that combined with eq.(\ref{eq:phio-linear}) gives:
\begin{equation}
\delta \phin = f \, \left[
\delta X_{\rm 14, c} 
+\delta \SNp
+\gammaNp\;  \delta \tc \right] + 
(1-f) \, \left[
\delta \carboneq 
+\delta \SCp 
+ \gammaCp \; \delta \tc
\right] 
\label{eq:phin-linear}
\end{equation}
where $f= \phio/\phin = 0.74$ is the ratio between $^{15}$O and $^{13}$N neutrino fluxes in SSMs \citep{Vinyoles:2017}.

\noindent{\em The abundance of carbon and nitrogen in the core of the  Sun}\\
Eqs. (\ref{eq:phio-linear}) and (\ref{eq:phin-linear}) describe the dependence of the CN-neutrino fluxes from the abundances of nitrogen $\nitrocen$ and carbon $\carboneq$ at the center of the Sun and close to $\rneq = 0.16R_\odot$, respectively.
These abundances are determined by the formation and  chemical evolution history of the Sun, i.e by the initial solar composition and by the subsequent action of nuclear reactions and elemental diffusion, as it is described in the following.
Let us first consider that the CN-cycle conserves the total number of CN-nuclei in the core of the Sun. This is shown in Fig.\ref{Fig:SSM} by the behaviour of the quantity:
\begin{equation} 
\Ncn \equiv X_{\rm 12}/12 + X_{\rm  13}/13 + X_{\rm 14}/14
\label{Ncn}
\end{equation} 
which is proportional to the total carbon+nitrogen number density ($X_{\rm  13}$ represents the carbon-13 mass abundance) and it is nearly constant in the solar core despite the action of nuclear reactions.
In the SSM paradigm, the radial dependence of $\Ncn$ is only due to elemental diffusion so that we can write:
\begin{equation}
\Ncn(r) = \Ncn_{\rm ini} \left[1 + \Dcn(r) \right]
\label{Dcn}
\end{equation}
where $\Ncn_{\rm ini}$ is the initial carbon+nitrogen abundance that is assumed to be uniform in the solar structure while the function $\Dcn(r)$ describes the effects of gravitational settling. It takes the value $\DcnC = 0.06$ at the center of the Sun that can be considered also representative for  $\rneq = 0.16R_\odot$, and $\DcnS = -0.09$ in external convective envelope according to SSM calculations \citep{Vinyoles:2017}.
It is useful to connect the core composition to photospheric abundances since these are observationally constrained by spectroscopic measurements. We thus write:
\begin{equation}
\NcnC = \NcnS \left[1 + \DcnCS \right]
\label{NcnCore}
\end{equation}
where $\NcnS$ ($\NcnC)$ is the carbon+nitrogen abundance in the external convective envelope (at the center) of the Sun while $\DcnCS = (\DcnC - \DcnS)/(1+\DcnS)=0.16$
represents the fractional difference between core and surface abundances.

The abundance $\nitrocen$ that controls the equilibrium production of CN-neutrinos is directly related to total abundance of carbon and nitrogen in the core of the Sun.
Indeed, for $r\lsim 0.1 R_\odot$ the CN-cycle is complete and all available carbon is essentially transformed into nitrogen, giving $\nitrocen \simeq 14\, \NcnC$ (see Fig.\ref{Fig:SSM}). We thus obtain the relation $\delta  \nitrocen  = \delta \NcnC$ that, by taking advantage of eqs.(\ref{Ncn},\ref{NcnCore}), can be rewritten as: 
\begin{equation}
\delta  \nitrocen   = a\; \delta X_{\rm 14,\, s}  + \left(1  -a\right) \;  \delta X_{\rm 12,\, s} + b \; \left(\DcnCS-0.16\right)
\label{deltaXn0}
\end{equation}
where $b = 1/(1+0.16) = 0.86$,  $a= 6\xi/(6\xi+7)\simeq 0.20$ and $\xi=(X_{\rm N,\,  s}/X_{\rm C, \, s}) \simeq 0.30$ is the surface nitrogen-to-carbon ratio in SSM. 
The first two terms of the r.h.s. in the above equation describe the effects produced by a variation of the surface composition. A modification of the chemical composition profile that is instead produced either "primordially" (e.g. by assuming that the Sun was not born chemical homogenous) or during the evolution (e.g. by anomalous diffusion) on time scales longer than carbon and nitrogen burning time at the solar center, is instead described in terms of a variation of $\DcnCS$ from the SSM value, i.e. by assuming $\DcnCS -0.016 \neq 0$.

A slightly more involved expression is obtained for the abundance $\carboneq$ that controls the non-equilibrium production of $^{13}{\rm N}$-neutrinos. In the relevant region $0.13 \lsim r/R_\odot \lsim 0.25$, the carbon-12 abundance differs from the surface value $X_{\rm 12,\, s}$ due to the action of elemental diffusion and $\CpS$ reaction only, since further reactions are inhibited by the $\NpS$ bottleneck. It can be approximately described as
\begin{equation}
\label{Xc1}
\carboneq \simeq X_{\rm 12,\, s} \left[1+\DcnCS \right] \exp\left( - \Dcpmedne \, t_{\odot} \right)
\end{equation}
where the quantity $\Dcpmed$ represents the carbon-12 burning rate
\begin{equation}
\Dcp  = \frac{\rho \, X}
{m_{u}} 
\svCp  
\label{D112}
\end{equation}
averaged over the Sun lifetime, see appendix for details. The maximal neutrino production is achieved at $\rneq \simeq 0.16\,R_\odot$ where the integrated burning rate is $\Dcpmedne \, t_{\odot} \simeq 1$. Indeed, in the inner core where $\overline{{\mathcal D}}_{112} \, t_{\odot} \gg 1$, carbon-12 abundance is too low to efficiently feed $\CpS$ reaction. On the other hand, the carbon-12 burning time is much larger than solar age (and thus $\CpS$ reaction is not effective) in more external regions where $\Dcpmed \ll (1/t_{\odot})$, as can be understood by considering that $ \Dcp \simeq \Dcpmed$. 
Taking this into account, we obtain the following relation:
\begin{equation}
\label{deltaXc1}
\delta X_{\rm 12}(r_{\rm ne}) =  \delta X_{\rm 12,\, s} + b \; \left(\DcnCS-0.16\right) - \delta \SCp - \gammaCp \, \delta \tc 
\end{equation}
where we considered that $\Dcpmedne \propto \SCp \,\tc^{\gammaCp} $.

\noindent{\em The final expressions the CN neutrino fluxes}\\
By using the above equations, we are able to calculate the dependence of neutrino fluxes produced in the CN-cycle on the properties of the Sun. 
By using eqs.(\ref{deltaXn0},\ref{deltaXc1}) into eqs.(\ref{eq:phio-linear},\ref{eq:phin-linear}), we obtain:
\begin{eqnarray}
\label{deltaphiCNFin}
\nonumber 
\delta \phio &=& \beta_{\rm O} \, \delta \tc   +  (1-a) \,  \delta X_{\rm 12,\, s} + a\, \delta X_{\rm 14,\, s} +  b \; \left(\DcnCS-0.16\right) +\delta  S_{114} \\
\delta \phin &=&  \beta_{\rm N} \, \delta \tc   +  (1-a') \,  \delta X_{\rm 12,\, s} + a'\, \delta X_{\rm 14,\, s} +  b \; \left(\DcnCS-0.16\right) +f\,\delta  S_{114}
\label{eq:deltaphicn}
\end{eqnarray}
with $\beta_{\rm O} = 20$, $f = 0.74$, $a=0.2$, $b =0.86$, $\beta_{\rm N} \equiv f \, \beta_{\rm O} = 15$ and $a' \equiv f\, a = 0.15$. Note that, in the derivation of the second equation, we took into account that the third and the fourth terms in the r.h.s of eq.(\ref{deltaXc1}) cancels the dependence of $\phin^{\rm (ne)}$ on $S_{112}$ and $T_{\rm c}$ expressed in eq.(\ref{eq:phin-ne-linear}). 
This is due to the fact that, as far as the $^{13}{\rm N}$-neutrino (non equilibrium) production rate is concerned, the effect of $\CpS$ cross section enhancement is compensated by the reduction of residual carbon-12 abundance due the more efficient carbon burning.  

\section{Numerical results and nuclear uncertainties}
\label{sec:numerical}
The expressions obtained for the neutrino fluxes can be compared with the results of SSMs calculations.
In particular, the numerical coefficients in eqs.(\ref{eq:deltaphipp},~\ref{eq:deltaphipep},~\ref{eq:deltaphibe},~\ref{eq:deltaphib},~\ref{eq:deltaphicn}) should reproduce the logarithmic derivatives of the neutrino fluxes with respect to the astrophysical factors of the relevant nuclear cross sections reported in Tab.~\ref{Tab:logderCS}. 
We see that a good agreement exists, indicating that all the major physical effects are included in our discussion and correctly described.
In the case of $\Spp$, we have to take into account that the role of this parameter is twofold; indeed, besides altering the efficiency of pp-reaction (at fixed temperature), this parameter also induces a variation of the central temperature of the Sun as described by Eq.(\ref{eq:deltatc-linear}).
This effect, combined with the strong temperature dependence of the fluxes, allow us to understand the large values for logarithmic derivatives reported in the first column of Tab.\ref{Tab:logderCS}. 

For completeness, we also discuss in the last two rows of Tab.~\ref{Tab:logderCS} the dependence of the helioseismic observable quantities $\ysur$ (surface helium abundance) and $\rcz$ (depth of the convective envelope) on nuclear reactions cross sections. We see that  $\Spp$ is the only nuclear parameter that affects the predictions for these quantities. The effects of $\Spp$ modifications on sound speed and density profiles are shown in the right panel of Fig.\ref{Fig:Spp}.
Finally, Tab.~\ref{Tab:logderOther} gives the logarithmic derivatives of neutrino fluxes and helioseismic quantities on other input parameters (beside nuclear cross sections) which are necessary to construct SSMs. These are:
%\begin{itemize}
%\item 
the solar age (\texttt{age}), luminosity (\texttt{lumi}) and the diffusion coefficients (\texttt{diffu});
%\item 
the opacity of solar plasma whose uncertainty is described in terms of two parameters $\kappa_{a}$ and $\kappa_{b}$ defined in Sect.~\ref{subsec:compo};
%\item 
the surface abundances of key elements
(C, N, O, Ne, Mg, Si, S, Ar, Fe) which are determined through spectroscopic measurements as discussed in Sect.~\ref{subsec:compo}.
%\end{itemize}
We can see that the logarithmic derivatives of the CN-neutrino fluxes with respect to the surface carbon and nitrogen abundances are correctly predicted by Eqs.~(\ref{eq:deltaphicn}).

The uncertainties in solar properties leading to environmental effects and chemical composition parameters, together with uncertainties in nuclear reaction cross sections propagate to SSM predictions which are affected by a {\em theoretical} (or {\em model}) error that can be estimated by Monte-Carlo techniques and/or linear propagation. By using this approach, the fractional error $\sigma_{Q}$ on a generic SSM prediction $Q$ can be obtained as the sum (in quadrature) of different contributions, according to:
\begin{equation}
\sigma^2_{Q} = \sum_{I} \left[ \alpha(Q,I) \right]^2 \sigma^2_I
\end{equation}
where $I=\texttt{age}, \texttt{lumi}, \dots$ indicates a specific input, $\sigma_{I}$ represents its fractional uncertainty and $\alpha(Q,I)\equiv d\ln Q/d\ln I$ is the logarithmic derivative of $Q$ with respect to $I$.
Tab.\ref{tab:uncert-param} contains the uncertainties $\sigma_{I}$ that have been considered for the construction of B16-SSMs (the surface composition errors are reported in Tab.\ref{tab:compo2}), see \cite{Vinyoles:2017} for details. By using these values, one is able to estimate the contribution $\delta Q_{I} \equiv \alpha(Q,I) \sigma_{I}$ of each input parameter to the total error budget of $Q$.
The dominant error sources for solar neutrino fluxes and helioseismic quantities are given in  Tab.\,\ref{tab:dom-err-sources}.\footnote{The total error due to opacity is obtained by combining in quadrature the contributions from $\kappa_a$ and $\kappa_b$.}

Focusing on nuclear reactions, we note that, despite the progress in the field, they are still an important uncertainty source for neutrino fluxes. In particular, the error contributions from $\Stq$ and $\Sbep$ are comparable to or larger than the uncertainties in the experimental determinations of $\phib$ and $\phibe$. 
As discussed in \cite{Vinyoles:2017}, the ability of solar neutrinos produced in the pp-chain to play a significant role in constraining physical conditions in the solar interior depends, although it is not the only factor, on pinning down errors of nuclear reaction rates to just $\sim 2\%$. For CN fluxes, we see that $\SNp$ is the dominant error source if composition is left aside. This is particularly relevant, especially in consideration of the fact that Borexino has just opened the era of CNO neutrino detection, obtaining for the first time $\sim 5\sigma$ direct experimental evidence for a non vanishing flux from the Sun \citep{Borexino:CNO}.  

For a correct evaluation of the importance of nuclear cross section, it should be remarked that, while neutrino fluxes generally change with variation in any of the input parameters, SSM predictions are strongly correlated with a single output parameter, the  core temperature $\tc$ \citep{Bahcall:1996, DeglInnocenti:1997, Haxton:2008, Serenelli:2013}. As a consequence, a  multi-dimensional set of variations of enviromental and chemical composition parameters $\{  \delta I  \}$  often collapses to a one-dimensional dependence on $\delta \tc$, where $\delta \tc$ is an implicit function of the variations $\{ \delta I \}$. The dominance  of $\tc$  as  the controlling  parameter for  neutrino fluxes can be exploited to cancel out uncertainties in the analysis of solar neutrino data. One can indeed form weighted ratios  $\Phi(\nu_{1})/\Phi(\nu_{2})^{x_{12}}$, or equivalently weighted fractional differences $\delta \Phi(\nu_{1}) - x_{12}\, \delta \Phi(\nu_{1})$ with respect to SSM predictions that are  nearly independent of $\tc$ and thus marginally affected by environmental effects and chemical composition, using the residual dependence on selected parameters to learn something about them. 

In \cite{Haxton:2008} and \cite{Serenelli:2013}, it was suggested to combine the CN-neutrino fluxes with the boron neutrino flux that, due to the exquisite precision of current experimental results and the large temperature sensitivity can be efficiently used as solar thermometer. As can be understood by considering Eqs.(\ref{eq:deltaphib}) and (\ref{eq:deltaphicn}), the following combinations can be formed: 
\begin{eqnarray}
\nonumber
\delta \phio  - x \, \delta \phib &=&  (1-a) \,  \delta X_{\rm 12,\, s} + a\, \delta X_{\rm 14,\, s} +  b \,\left(\DcnCS-0.16\right) \\
\label{eq:deltaphiob}
&+&\delta  S_{114} - x \left(\frac{\delta S_{11}}{2} - \frac{\delta S_{33}}{2} +\delta S_{34} + \delta S_{17} -\delta S_{e7} \right)\\
\nonumber
\delta \phin  - x' \, \delta \phib &=&  (1-a') \,  \delta X_{\rm 12,\, s} + a'\, \delta X_{\rm 14,\, s} +  b \,\left(\DcnCS-0.16\right) \\
&+&\delta  S_{114} - x' \left(\frac{\delta S_{11}}{2} - \frac{\delta S_{33}}{2} +\delta S_{34} + \delta S_{17} -\delta S_{e7} \right)
\label{eq:deltaphinb}
\end{eqnarray}
where $x = \beta_{\rm O}/\beta_{\rm B} \simeq 0.8$ and $x' = f\, x \simeq 0.6$, that are independent from $\delta \tc$. This possibility is extremely important because it allows us to cancels out the dependence on the radiative opacity (implicit in $\delta\tc$). The uncertainty of available opacity calculations is indeed not easily quantified and may be potentially underestimated. Moreover, it breaks the degeneracy between composition and opacity effects on solar observable properties. 
Indeed, the considered flux combinations only depend on the carbon and nitrogen abundance in the solar core allowing us to test the chemical composition and evolution of the Sun. The first two terms in the r.h.s. of Eqs.~(\ref{eq:deltaphiob},\ref{eq:deltaphinb}) quantify the effects of a variation of the surface C and N abundances. A change of the diffusion efficiency is instead described in terms of a variation of $\DcnCS$ from the SSM value, i.e. by assuming $\DcnCS -0.016 \neq 0$.
It should be remarked that the ability to probe solar composition by using this approach is only limited by experimental accuracy of flux determinations and by nuclear cross section uncertainties. 

While the above relationships are based on the simplified arguments discussed in the previous section, the optimal combinations $\delta \Phi(\nu_1) - x_{12} \delta \Phi(\nu_2)$, or equivalently weighted ratios  $\Phi(\nu_{1})/\Phi(\nu_{2})^{x_{12}}$, can be  determined by using the power-law coefficients from \cite{Vinyoles:2017} given in Tab.\ref{Tab:logderOther}. The parameter $x_{12}$ is obtained by minimizing the residual 
\begin{equation}
\rho=\sum_{I=1}^{N}\left[\alpha(\nu_1,I)-x_{12} \, \alpha(\nu_1,I) \right]^2 \sigma_I^2
\end{equation}
where the sum extends to the $N$ input parameters whose dependence we want to cancel out and $\sigma_I$ are the corresponding uncertainties. The minimal value for $\rho$ gives the intrinsic error in the considered approach.
This method, originally proposed by \cite{Haxton:2008} and \cite{Serenelli:2013}, has been recently adapted to Borexino in \cite{Agostini:2020lci}.
By taking into account that the measured CNO neutrino signal in Borexino is basically probing $\delta \phi^{\rm BX}_{\rm CNO} \equiv \xi\, \delta \phio + (1-\xi)\, \delta \phin$ with $\xi = 0.764$ , it was concluded that the surface composition of the Sun can be probed by the combination:
\begin{eqnarray}
\nonumber
\delta R^{\rm BX}_{\rm CNO} - 0.716 \, \delta \phib &=& 0.814 \, \delta X_{\rm 12,\, s} + 0.191 \, \delta X_{\rm 14,\, s} \\ 
%&\pm& 2\%\, (^{8}{\rm B}) 
&\pm& 0.5\%\, {\rm (env)}
\pm 9.1\%\,{\rm (nucl)} \pm  2.8\%\,{\rm (diff)}
\label{eq:CN-core}
\end{eqnarray}
where $\delta R^{\rm BX}_{\rm CNO}$ is the fractional difference of the observed CNO signal with respect to SSM expectations and the quoted uncertainties are obtained by propagating errors of SSM input parameters. 
The error budget is presently dominated by the uncertainty of the CNO signal Borexino measurement.
However, a relevant error ($\sim 10\%$) is also provided by nuclear reactions, with the largest contributions coming from $\SNp$ (7.6\%), $\Stq$ (3.4\%), and $\Sbep$ (3.5\%).
In the perspective of future improvements of the CNO signal determination, it is evidently important to have reliable and accurate determinations of these cross sections.

 % ============================
\begin{table}[!ht]
\begin{center}{
\footnotesize
\begin{tabular}{l|cccccccc} 
\hline \hline
       & $S_{11}$  &  $ S_{33}$    & $S_{34}$  &   $S_{e7}$ &    $S_{17}$ &   $S_{hep}$ &  $ S_{114}$  &  $S_{116}$   \\
\hline 
$\phipp$   & 0.101 &  0.034 &  -0.066 &   0.000 &  0.000 &  0.000 & -0.006 & -0.000  \\ 
$\phipep$ & -0.222 &  0.049 & -0.095 &  0.000 &  0.000  & 0.000 & -0.010 &  0.000  \\
$\phihep$ & -0.104 & -0.463 & -0.081 &  0.000 &  0.000 &  1.000 & -0.006 & -0.000  \\
$\phibe$ & -1.035  &-0.440 &  0.874 &  0.002 & -0.001 &  0.000 & -0.001 &  0.000   \\
$\phib$  & -2.665 & -0.419 &  0.831 & -0.998  & 1.028 &  0.000 &  0.007 &  0.000   \\
$\phin$  &-2.114 &  0.030 & -0.061 &  0.001 &  0.000 &  0.000 &  0.762 &  0.001   \\
$\phio$  &-2.916 &  0.023 & -0.050  & 0.001 &  0.000 &  0.000 &  1.051 &  0.001   \\
$\phif$  & -3.072 &  0.021 & -0.046 &  0.001 &  0.000 &  0.000 &  0.007 &  1.158   \\
\hline
$\ysur$   &  0.131 & -0.005 &  0.010 &  0.000  & 0.000 &  0.000 &  0.001 &  0.000  \\
$\rcz$  & -0.059 &  0.002 & -0.004 &  0.000 &  0.000 &  0.000 & 0.000 & 0.000 \\
\hline
\end{tabular}
}\end{center}
\vspace{-0.5cm} 
\caption{{\small\em The logarithmic derivatives $\alphaQI$ of the solar neutrino fluxes with respect to nuclear input parameters calculated in B16-HZ SSMs.
\label{Tab:logderCS} } }
\vspace{0.5cm}
\end{table}

%=============================
\begin{table}[!ht]
\begin{center}{
\footnotesize
\begin{tabular}{l|ccccc|ccccccccc} 
\hline \hline
      & \texttt{age} &   \texttt{diffu}  &    \texttt{lumi}  & \texttt{$\kappa_a$} &  \texttt{$\kappa_b$}
      &     C  &     N  &     O   &   Ne  &    Mg  &    Si    &   S   &  Ar  &    Fe  \\
\hline
$\phipp$ &-0.085 & -0.013  & 0.773 & -0.084 & -0.019
& -0.007 & -0.001 & -0.005 & -0.005 & -0.003 & -0.009 & -0.006 & -0.001 & -0.019 \\  
$\phipep$ & -0.003 & -0.018 &  0.999 & -0.270  &-0.001 &  -0.014 & -0.002  &-0.011 & -0.005&  -0.003&  -0.012 & -0.013 & -0.004  &-0.060 \\ 
$\phihep$ &-0.125 &  -0.039 &  0.149 & -0.395  &-0.107 &-0.008 & -0.002  &-0.024 & -0.018 & -0.016 & -0.036 & -0.027 & -0.006 & -0.066 \\ 
$\phibe$ & 0.753 &  0.132 &   3.466  & 1.332 &  0.380 &-0.000&   0.002&   0.057 &  0.053 &  0.052 &  0.106 &  0.075 &  0.018 &  0.209  \\
$\phib$ & 1.319  & 0.278 &  6.966  & 2.863 &  0.658 &  0.022 &  0.007 &  0.128&  0.102&   0.092&  0.198 &  0.138  & 0.034 &  0.498   \\
$\phin$ & 0.863 &  0.345  & 4.446  & 1.592 &  0.314
&  0.864 &  0.154  & 0.073 &  0.051 &  0.047 &  0.110 & 0.078  & 0.020 &  0.272  \\
$\phio$ & 1.328 &  0.395 &  5.960   & 2.220 &  0.456
&  0.819 &  0.209 &  0.104 &  0.075 &  0.068 &  0.153&   0.107 &  0.027 &  0.388  \\
$\phif$ & 1.424 &  0.418 &  6.401  & 2.427 &  0.503 & 0.026 &  0.007&  1.112 &  0.082 &  0.074  & 0.167 &  0.116 &  0.029  & 0.424  \\
\hline
$\ysur$ & -0.195 & -0.077 &  0.351 &  0.608 &  0.255
  &  -0.008 & -0.001&  0.019 &  0.032 & 0.032 &  0.062  & 0.042 &  0.010 &  0.084   \\
$\rcz$ & -0.081 & -0.018 & -0.016 &  0.008 & -0.079
& -0.003 & -0.003 & -0.024  &-0.012 & -0.004 &  0.003 &  0.005 &  0.001&  -0.008  \\
\hline
\end{tabular}
}\end{center}
\vspace{-0.5cm} 
\caption{{\small\em The logarithmic derivatives $\alphaQI$ of the solar neutrino fluxes with respect to solar properties that produce environmental effects and chemical composition parameters calculated in B16-HZ SSMs. 
\label{Tab:logderOther} } }
\vspace{0.5cm}
\end{table}

%=============================
\begin{table}[t]
\begin{center}{
\footnotesize
\begin{tabular}{l|ccccc|ccccccccc} 
\hline \hline 
&  \texttt{Age} & \texttt{Diffu} & \texttt{Lum} &  \texttt{$\kappa_a$} &  \texttt{$\kappa_b$} & $S_{11}$ & $S_{33}$ &  $S_{34}$ & $S_{17}$ & $S_{e7}$ & $S_{114}$ & $S_{116}$ & $S_{\rm hep}$ \\ 
\hline
&  0.0044 & 0.15 & 0.004 &  0.02 & 0.067  &
0.01 & 0.052  & 0.052  & 0.047 & 0.02 & 0.075 & 0.076   & 0.30 \\
\hline
\end{tabular}
}\end{center}
\vspace{-0.5cm} 
\caption{{\small\em The fractional uncertainties of enviromental and nuclear input parameters in SSM construction. 
\label{tab:uncert-param} } }
\vspace{0.5cm}
\end{table}

\begin{table}
\footnotesize
\begin{center}
\begin{tabular}{l | l@{\hskip 0.1cm}r@{\hskip 0.6cm} l@{\hskip 0.1cm}r@{\hskip 0.6cm} l@{\hskip 0.1cm}r@{\hskip 0.6cm} l@{\hskip 0.1cm}r }
\hline
\hline
Quant.&\multicolumn{8}{c}{Dominant theoretical error sources in \%} \\
\hline
$\phipp$ & $\lsun$:& 0.3 & $S_{34}$: &  0.3 & $\kappa$: & 0.2 & Diff: & 0.2  \\
$\phipep$ & $\kappa$: &0.5 & $\lsun$: & 0.4 & $S_{34}$: & 0.4 & $S_{11}$: & 0.2 \\
$\phihep$& $S_{\rm hep}$: &30.2 & $S_{33}$: & 2.4 & $\kappa$: & 1.1 & Diff: & 0.5\\
$\phibe$& $S_{34}$: &4.1 & $\kappa$: & 3.8 & $S_{33}$: & 2.3 & Diff: & 1.9 \\
$\phib$& $\kappa$: &7.3 & $S_{17}$: & 4.8 & Diff: & 4.0 & $S_{34}$: & 3.9\\
$\phin$& C: &10.0 & $S_{114}$: & 5.4 & Diff: & 4.8 & $\kappa$: & 3.9\\
$\phio$& C: &9.4& $S_{114}$: & 7.9 & Diff: & 5.6 & $\kappa$: & 5.5 \\
$\phif$& O: &12.6 & $S_{116}$: & 8.8 & $\kappa$: & 6.0 & Diff: & 6.0  \\
\hline
%$\alphamlt$& O: &1.3 &  Diff: & 1.2 & $\kappa$: & 0.7  & Ne: & 0.7 \\
%$\yini$&  $\kappa$: &1.9 & Ne: & 0.5 & Diff: & 0.4 & Ar: & 0.3 \\
%$\zini$&  O: &4.7 & C: & 2.0 & Ne: & 1.7 & Diff: & 1.6 \\
$\ysur$& $\kappa$: &2.2 & Diff: & 1.1 & Ne: & 0.6 & O: & 0.3 \\
%$\zsur$&  O: &4.8 & C: & 2.0  & Ne: & 1.8 & $\kappa$: & 0.7 \\
$\rcz$&  $\kappa$: &0.6 &  O: & 0.3  & Diff: & 0.3 & Ne: & 0.2 \\
\hline
\end{tabular}
\end{center}
\caption{\em\small Dominant theoretical error sources for neutrino fluxes and the main characteristics of the SSM. 
\label{tab:dom-err-sources}}
\end{table}

\section{Concluding remarks}
\label{sec:conclusions}
A fundamental part in solar model calculations is the knowledge of the rates of nuclear reactions involved in the generation of solar nuclear energy.
During the last decades, we experienced a substantial progress in the accuracy of SSM calculations that was made possible, among the other ingredients, by the continuous improvements of nuclear cross sections that are now typically determined with $\sim 5\%$ accuracy.
However, SSMs have now to challenge new puzzles, like e.g. the solar composition problems. Moreover, SSM neutrino flux predictions, which are directly affected by nuclear cross sections uncertainties, have to be compared against very accurate observational determinations, having errors at ${\rm few} \%$ level or better e.g. for $\phibe$ and $\phib$.

As a consequence, further work is needed on the side of nuclear reactions. Indeed, nuclear uncertainties have a non negligible role in SSMs error budget.
As an example, the error contributions from $\Stq$ and $\Sbep$ are about a factor 2 larger than the
uncertainties in the experimental determinations of  $\phibe$ and $\phib$.
As it is discussed in \cite{Vinyoles:2017}, the few percent systematics in the determination of these reaction rates is still a relevant source of difficulty in using neutrino fluxes as constraints to solar model properties.
The astrophysical factor $\SNp$ is morevoer a relevant error source for CN neutrino fluxes. 
This last point is particularly important after Borexino opened the era of CNO neutrino detection, obtaining the first ever direct evidence of a non vanishing CN neutrino signal from the Sun.
In the perspective of future and more accurate measurements, nuclear uncertainties can become a limiting factor in the possibility to use the CN-neutrinos, in combination with $^8{\rm B}$ neutrinos, to directly probe the solar composition, thus addressing the solar composition problem. 
%
%In conclusion, it would be desirable to further improve our knowledge of nuclear cross sections, in particular for $\hetheqS$, $\bepS$ and $\NpS$ reactions, with the ambitious goal of reducing their uncertainties by a factor $\sim 2$, if possible, in the next future.
%
At the moment, the nuclear error contribution to CN-core abundance uncertainty is $\sim 10\%$, see eq.~(\ref{eq:CN-core}).
This is comparable to the error in CN-surface abundance determinations (0.05 dex in LZ composition) and only a factor $\sim 2$ smaller than the difference between HZ and LZ results, which can be regarded as an estimate of the systematic shift in the surface abundances produced by advances in stellar spectroscopy during the last 20 years.
We remark that a high accuracy determination of the solar core composition could be used not only to discriminate among different solar surface admixtures but also to test the  chemical evolution scheme employed by SSMs, e.g. by verifying the effect of elemental diffusion according to which core abundances are expected to be $\sim 15\%$ larger than surface values.

In conclusion, it would be desirable to further improve our knowledge of nuclear cross sections, in particular for $\hetheqS$, $\bepS$ and $\NpS$ reactions.
As we discussed in the introduction, the history of SSMs appears to be formed by three large chapters, during which the knowledge of nuclear rates improved at each stage by about a factor two with respect to the previous period, up to the present situation in which the leading cross section in pp-chain and CN-cycle are typically determined with $\sim 5\%$ accuracy.
The ambitious goal for the next stage could be a further  factor $\sim 2$ reduction, in such a way that nuclear reactions uncertainties will not represent a limiting factor in constraining the physical conditions of solar interior.

\section*{Acknowledgements}
FV acknowledges support by ‘Neutrino and Astroparticle Theory Network’ under the programme PRIN 2017 funded by the Italian Ministry of Education, University and Research (MIUR) and INFN Iniziativa Specifica TAsP. AS acknowledges support by the Spanish Government through the MICINN grant PRPPID2019-108709GB-I00 and by the COST action ChETEC CA16117.

\bibliographystyle{frontiersinSCNS_ENG_HUMS} % for Science, Engineering and Humanities and Social Sciences articles, for Humanities and Social Sciences articles please include page numbers in the in-text citations
%\bibliographystyle{frontiersinHLTH&FPHY} % for Health, Physics and Mathematics articles
%\bibliography{test}

%%% Make sure to upload the bib file along with the tex file and PDF
%%% Please see the test.bib file for some examples of references

\newpage

\section*{Appendix: The $^{12}$C abundance in non-equilibrium region}

In the region $0.13\lsim r/R_\odot \lsim 0.25 $, the CN-cycle is incomplete; carbon-12 is partially burned by $\CpS$ while nitrogen-14 is not effectively processed by $\NpS$ reaction. 
If we neglect elemental diffusion, the equation that describes the time evolution of carbon-12 is (in lagrangian coordinates):
\begin{equation}
\label{c-evol}
\frac{\partial X_{\rm 12}}{\partial t} = 
- X_{\rm 12}\, {\mathcal D}_{112}
\end{equation}
where the carbon-12 burning rate ${\mathcal D}_{112}$ is given by:
\begin{equation}
{\mathcal D}_{112}  = \frac{\rho \, X}
{m_{u}} 
\langle \sigma v \rangle_{112}  
\label{D112App}
\end{equation} 
The solution of Eq.~(\ref{c-evol}) is:
\begin{equation}
\label{XcA}
X_{\rm 12} =  
X_{\rm 12, ini}\, \exp\left(- {\overline{\mathcal    D}}_{112} \; t_{\odot} \right)
\end{equation}
where $X_{\rm 12, ini}$ is the initial abundance and ${\overline{\mathcal    D}}_{112}$ is given by:
\begin{equation}
{\overline{\mathcal    D}}_{112} \equiv \frac{1}{t_\odot} \,
\int^{t_\odot}_{0} dt \; {\mathcal D}_{112} \,.
\label{D112int}
\end{equation}
We include {\em a-posteriori} the effecs of elemental diffusion by
replacing $X_{\rm 12, ini} \to X_{\rm 12, ini} (1+\Delta(r))$ with the
function $\Delta(r)$ defined in Eq.(\ref{Dcn}).
We can then recast in terms of the surface carbon abundance, obtaining:
\begin{equation}
\label{XcB}
X_{\rm 12} =  
X_{\rm 12,\, s} \, \left[ 1+\Delta^{\rm (cs)} \right]\,
\exp\left(- {\overline{\mathcal    D}}_{112} \; t_{\odot} \right).
\end{equation}
where $\Delta^{\rm (cs)}=0.16$ is the fractional difference between core and surface abundances induced by elemental diffusion.

%\end{appendix}

%%% If you are submitting a figure with subfigures please combine these into one image file with part labels integrated.
%%% If you don't add the figures in the LaTeX files, please upload them when submitting the article.
%%% Frontiers will add the figures at the end of the provisional pdf automatically
%%% The use of LaTeX coding to draw Diagrams/Figures/Structures should be avoided. They should be external callouts including graphics.

\end{document}